\documentclass{article}
\usepackage{graphicx}
\usepackage{amsmath}
\usepackage{amsbsy}
\usepackage{epstopdf, epsfig}
\usepackage[a4paper, total={6in, 8in}]{geometry}
\usepackage{xspace}
\newcommand{\co}{CO\ensuremath{_\textrm{2}}\xspace}
\newcommand{\pcs}{\ensuremath{P_\textrm{c}}-\ensuremath{ S}\xspace}
\newcommand{\pcc}{\ensuremath{P_\textrm{c}}-\ensuremath{\overline{\chi}}\xspace}
\newcommand{\pcsc}{\ensuremath{P_\textrm{c}}-\ensuremath{ S}\xspace -\ensuremath{\overline{\chi}}\xspace}

\title{Impact of Time-Dependent Wettability Alteration on the Dynamics of Capillary Pressure}

\author{Abay M. Kassa$^{1,2}$ \and
  Sarah E. Gasda$^{2}$ \and
  K. Kumar$^{1}$ \and
 \and F. A. Radu$^{1}$}
\date{}
\begin{document}
\maketitle
\noindent ${}^1$ Department of Mathematics, University of Bergen, P. O. Box 7800, 5020 Bergen, Norway.\\[5pt]
${}^2$ NORCE, Nyg{\aa}rdsgaten 112, 5008 Bergen, Norway.\\[5pt]
Corresponding author: Abay M. Kassa (E-mail: abka@norceresearch.no)
\begin{abstract}
Wettability is a pore-scale property that has an important impact on capillarity, residual trapping, and hysteresis in porous media systems. In many applications, the wettability of the rock surface is assumed to be constant in time and uniform in space. However, many fluids are capable of altering the wettability of rock surfaces permanently and dynamically in time. Experiments have shown wettability alteration can significantly decrease capillarity in \co storage applications. For these systems, the standard capillary-pressure model that assumes static wettability is insufficient to describe the physics. In this paper, we develop a new dynamic capillary-pressure model that takes into account changes in wettability at the pore-level by adding a dynamic term to the standard capillary pressure function. We simulate the dynamic system using a bundle-of-tubes (BoT) approach, where a mechanistic model for time-dependent contact angle change is introduced at the pore scale. The resulting capillary pressure curves are then used to quantify the dynamic component of the capillary pressure function. This study shows the importance of time-dependent wettability for determining capillary pressure over timescales of months to years. The impact of wettability has implications for experimental methodology as well as macroscale simulation of wettability-altering fluids.
\end{abstract}

\section{Introduction}
Wettability plays an important role in many industrial applications, in particular subsurface porous media applications such as enhanced oil recovery (EOR) and \co storage \cite{Bonn09, Iglauer2016,  Yu2008, Iglauer2014, RefBlunt}. The wetting property of a given multiphase system in porous media is defined by the contact angle, which describes the affinity of one fluid for the surface of the medium over another. The contact angle is determined from surface chemistry and associated forces acting at the molecular scale along the fluid-fluid-solid interface \cite{Bonn09}. In porous media applications,  
microscale wettability determines the strength of pore-scale capillary forces and the movement of interfaces between individual pores. At the macroscale, wettability impacts upscaled quantities and constitutive functions such as residual saturation, relative permeability, and capillary pressure, which in turn affect large-scale multiphase flow behavior. 

The standard assumption is that wettability is a static property of the multiphase system. However, the composition of many fluids is capable of provoking the surfaces within pores to undergo wettability alteration (WA) via a change in contact angle.  
A change in contact angle can alter capillary forces at the pore scale, and thus affect residual saturations of the system \cite{Ahmed2001, RefBlunt97}. This effect has been exploited extensively in the petroleum industry, where optimal wetting conditions in the reservoir are obtained through a variety of means that includes chemical treatment, foams, surfactants and low-salinity water flooding (see for example \cite{Morrow86, Buckley1998, Jabhunandan95, RefHaag17, RefSingh16}). 

Wettability is also recognized as a critical factor in geological \co sequestration which exerts an important role on caprock performance \cite{RefKim12, Tokunaga13}. WA can influence the ability to prevent \co leakage from the reservoir by altering the capillary forces that act to trap \co as buoyant phase beneath a low-permeability caprock. This process is highly dependent on \co being a strongly non-wetting fluid, and WA may lead to conditions that allow for \co leakage \cite{Chiquet07a,  Chaulbud2009}.  In addition, WA can affect residual saturation, results in an impact on trapping efficiency of injected \co \cite{Iglauer2014}. Therefore, reliable quantification of wettability is key to understand the macroscale processes in subsurface applications.

Despite the fact that WA is known to impact macroscale capillarity and relative permeability behavior, few detailed measurements are available to characterize the alteration of the constitutive function themselves. 
Plug et al., \cite{Plug07} have reported brine-\co (gas, liquid) drainage-imbibition experiment and showed capillary instability for supercritical \co-brine system, meaning that the capillary pressure measurements changes steadily over time. The imbibition curve also exhibited a significant deviation from the expected curve that was not explained by classical scaling arguments. The authors proposed WA as an explanation but did not explicitly measure any changes in contact angle. Additionally, recent experiments measured capillary pressure curves for a silicate sample using a fluid pairing of supercritical \co and brine \cite{RefW}. Repeated drainage-imbibition cycles were performed over a period of 6 months, and a clear reduction in capillary pressure was recorded for each subsequent drainage cycle. The authors also attributed these deviations from the expected capillary curve to a change in wettability of the rock sample over time due to \co exposure, similar to aging. This hypothesis was confirmed through observations of a wetting angle increase from $0^\circ$ to $75^\circ$ after 6-months exposure. It is also reported similar \pcs instability and deviation in dolomite/carbonate \cite{RefWang}, and quartz \cite{Tokunaga2013, RefWang} sands for sc\co -brine system. More literature on WA and \pcs measurements can be found in \cite{Tokunaga13}. 

The above experiments reveal that capillary pressure curves are not static when significant WA occurs, despite the fact they were performed following the standard multi-step outflow procedure, i.e. where ``equilibrium''  is obtained after each incremental step in pressure. Therefore, the standard equilibrium capillary pressure models cannot be readily applied without additional dynamics to capture the impact of WA. Capillary pressure dynamics due to WA is fundamentally different from the well known non-equilibrium flow dynamics (for example see \cite{RefHassanizadeh, RefDahle}) and contact angle hysteresis, i.e. receding and advancing contact angle \cite{Krumpfer10, Eral12} because these dynamics disappear when the system is at equilibrium.  

In petroleum reservoir simulation, there have been some theoretical investigations that assume an instantaneous WA to model the constitutive relations and immiscible flow simulations \cite{Delshad09, RefLashgari, Yu2008, Andersen15, Adibhatla05}. The WA mechanism is incorporated by interpolation between the final and initial wetting-state constitutive relations \cite{Andersen15,Delshad09}.  Lashgari and co-authors \cite{RefLashgari} have derived an instantaneous WA model from Gibbs and Langmuir adsorption models. The proposed WA model is coupled to \pcs and relative permeability curves through residual saturation. These models neglect exposure time dependency of WA and also the impact of pore-scale heterogeneity in wetting properties in the macroscale simulations.
In the available literature, only one study \cite{RefAl-Mutairi} has included the effect of exposure time on WA and constitutive relations for core scale simulation. But this numerical study does not sufficiently incorporate or upscale pore-scale processes to macroscale constitutive laws.
 
To our knowledge, a rigorous mathematical characterization of dynamics in \pcs functions introduced by exposure to a WA agent has not been previously performed. In this paper, we develop a new model for dynamic capillary pressure that is derived using direct simulation of \pcs curves from a pore-scale model. WA is introduced at the pore scale using a mechanistic model for contact angle change as a function of exposure time to a WA fluid. This mechanistic model is not intended as a replacement for detailed surface chemistry models, but is used merely as a convenient mathematical way to represent changing contact angle in time. We acknowledge that the field of wettability alteration at the micro- and nano-scales is an active area of research covering a broad range of complexity. We emphasize that this study does not directly address wettability alteration at pore surfaces, which is well beyond the scope of this paper.

The pore-scale flow model is represented by a cylindrical bundle-of-tubes model. This pore-scale approach is simplistic in that it does not represent interactions between pores or realistic porous geometry (i.e. converging-diverging throat diameters). Additionally, there is no natural hysteresis or residual saturation with cylindrical tubes. However, a simple pore-scale model is sufficient as a first step towards develop time-dependent correlated expressions for \pcs. It is also important to investigate the impact of wettability alteration apart from other complicating processes that can occur in more realistic pore-network models.  
\section{Approach} 
\label{sec:approach}
The extended $P_c$-$S_w$ relationship introduces a dynamic component that captures the changing wettability as measured by the deviation of the dynamic capillary pressure from the equilibrium (static) capillary pressure. This relationship can be described as follows:
\begin{equation}
    P_c(\cdot)-P_c^{\rm st,i}:=f^{\rm dyn}(\cdot),
    \label{eq:dynPc}
\end{equation}
where $P_c^{\rm st,i}$ represents the capillary pressure for the system given a static initial wetting state, while $f^{\rm dyn}$ represents the deviation from the static state. The static curve can be characterized by a number of well-known capillary pressure models, such as van Genuchten or Brooks-Corey. 
Static systems studied here are  described by the Brooks-Corey model,
\begin{equation}\label{eq:bc1}
P_c^{\rm st} = {c_w}\left(\frac{S_w-S_{wc}}{1-S_{wc}}\right)^{-a_w},
\end{equation}
where $c_w$ is the entry pressure, $1/a_w$ is the pore-size distribution index, whereas $S_{wc}$ is the residual water saturation. The static curve is equivalent to the equilibrium capillary pressure for systems that do not undergo wettability alteration. 

The objective of this study is to characterize and quantify the dynamic term $f^{\rm dyn}$, the key term of interest in the \pcs model, for a system that undergoes wettability alteration. We propose an interpolation model can be applied for this system, where the dynamic term acts to shift the capillary pressure between two end wetting states in time. This type of model has been used in petroleum reservoir simulation  \cite{Andersen15, Delshad09, Yu2008,RefLashgari}. To obtain an interpolation model, the dynamic component in Equation (\ref{eq:dynPc}) can be scaled by the difference between two static curves, each representing the initial and final wetting-state capillary pressure curves, to give a non-dimensional quantity we call the \emph{dynamic coefficient}, which is defined as follows,
\begin{equation}\label{eq:dynPc2}
\omega\big(P_c^{\rm st,f}-P_c^{\rm st,i}\big) = f^{\rm dyn},
\end{equation}
where $P_c^{\rm st,f}$ is the final wetting state capillary pressure. This can be substituted into Equation (\ref{eq:dynPc}) to obtain a dynamic interpolation model for capillary pressure,
\begin{equation}
    P_c=(1-\omega)P_c^{\rm st,i} + \omega P_c^{\rm st,f},
    \label{eq:dynPc_interp}
\end{equation}
We note that in the model presented above, we have defined the ``total'' capillary pressure $P_c$ as simply the measured difference in phase pressures at any point in time. In a reservoir simulation, this would be capillary pressure in a given grid cell, whereas in a multi-step outflow experiment, it corresponds to the pressure drop across the experimental cell. For quasi-static displacement in a bundle of capillary tubes, $P_c$ is the difference in reservoir pressures, i.e., $P_c = P_l^{res}-P_r^{res}$, (see Figure \ref{Bundfigure}). 
%
%
The exact nature of $\omega$ and its functional dependencies can only be determined from a full characterization of \pcs curves under different conditions. These curves can be derived from laboratory experiments, but this approach is costly and time-consuming. Alternatively, one may take a more theoretical approach by simulating \pcs curves using a pore-scale model that includes the impact of WA.

For a system that undergoes WA, a significant change in contact angle could lead to the wetting phase becoming non-wetting and vice versa. For clarity, we will continue to use $w$ subscript for the phase that was originally wetting and $nw$ for the phase that is originally non-wetting regardless of the actual state of wettability in the system.
\subsection{Capillary flow in cylindrical tubes}\label{sec:11}
There are various choices of pore-scale models available. The easiest to implement and analyze is the bundle of tubes (BoT) model which is a collection of capillary tubes with a distribution of radii. Herein, we describe the model and the approach for implementation of time-dependent WA within individual capillary tube. 

The BoT model has been applied in many immiscible multiphase flow studies and is employed to study constitutive relations and fluid dynamics in general \cite{RefB, RefDahle, RefHelland}. 
The BoT model is a popular approach due to the simplicity of implementation and the ability to study the balance of energy and forces directly at the scale of interfaces. The average behaviour of a BoT model can then be used to inform better constitutive functions at the macroscale. However, direct use of simulated capillary or relative permeabilities curves is usually not advised. This is because a collection of capillary tubes is not representative of the interacting nature of pores and pore throats. The geometry is over-simplified with respect to converging and diverging pore throats. Some of the shortcomings of BoT models can be addressed through interactive BoT models \cite{RefDong}, or tubes of more complex geometry. These more advanced BoT models are beyond the scope of this paper. 

\begin{figure}[t!]
\center
\includegraphics[width=0.75\textwidth]{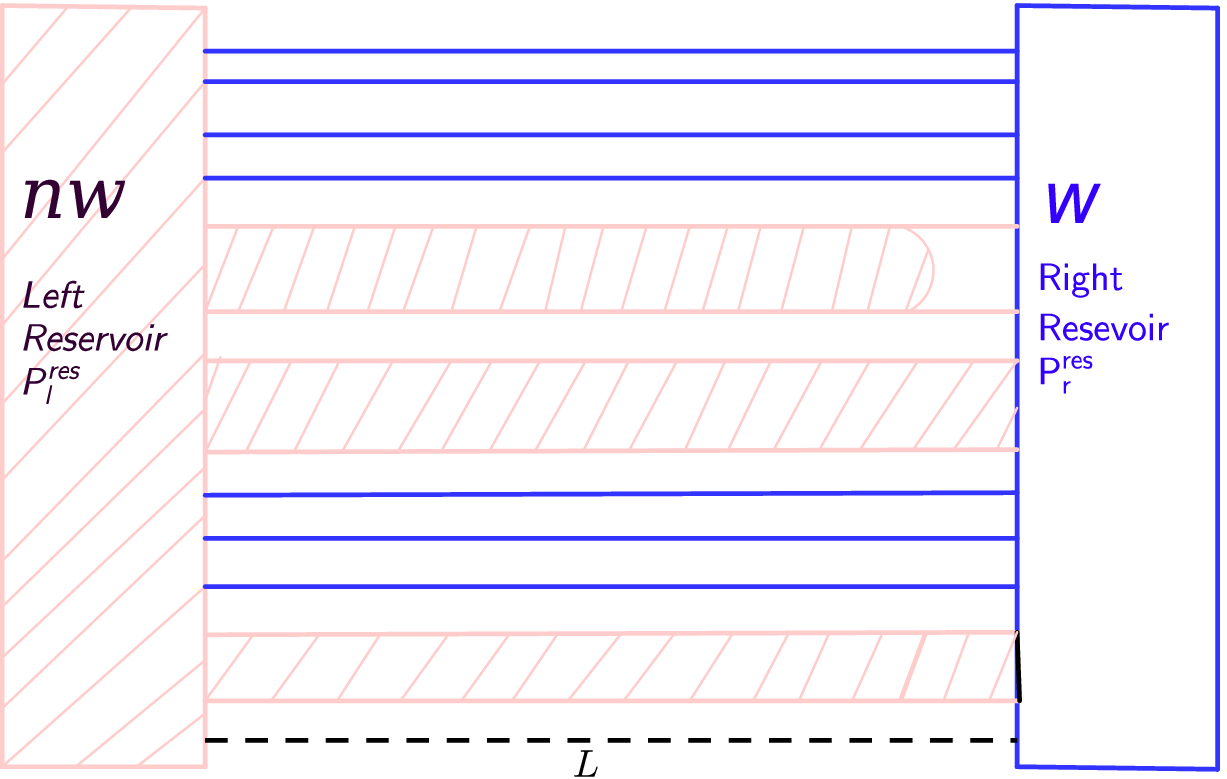}%
\caption{Fluid displacement in non-interactive bundle of tubes (BoT). Here, the left reservoir contains non-wetting fluid that displaces the wetting fluid to the right and vice versa.}\label{Bundfigure}
\end{figure}

In this paper, we consider cylindrical bundle of tubes having length, L, and is shown in Figure \ref{Bundfigure}.
These tubes are designed to connect wetting (right) and non-wetting (left) reservoirs with pressures labeled as $P_{\rm r}^{\rm res.}$  and $P_{l}^{\rm res.}$ respectively. Let the reservoir pressures difference be defined as:
\begin{equation}\label{rpd}
\Delta P = P^{res}_{l}-P^{res}_{ r}.
\end{equation}
Equation (\ref{rpd}) indicates that the pressure drop across the bundle is equal to the measured capillary pressure $P_c$ defined previously in Equation (\ref{eq:dynPc}). 

We consider the tubes to be  filled with water and are water-wet initially. To displace the water phase, the pressure drop has to exceed the entry capillary pressure \cite{RefDahle}:
\begin{equation}\label{rpd2}
\Delta P > P_c^{\rm mic}(R_m, \theta_m),
\end{equation}
where $P_c^{\rm mic}(R_m, \theta_m)$ is given by the Young's equation %
\begin{equation}\label{rpdc}
 P_c^{\rm mic}(R_m, \theta_m) = \frac{2\sigma_{ow} \cos(\theta_m)}{R_m}, ~ m = 1,2,\cdots,N,
\end{equation}
where $R_m$ is the $m$-th tube radius, $N$ stands for number of tubes, $\sigma$ is fluid-fluid interfacial tension (IFT) and $\theta_m$ measures the wetting property of the rock surface (it may be static or dynamic in time per the $m$-th tube). As long as condition (\ref{rpd2}) is satisfied, the fluid movement across the length of a given tube can be approximated by the  Washburn flow model, 
\begin{equation}\label{eqbundle3}
q_m =\frac{R_m^2(\Delta P - P_c^{\rm mic}(R_m, \theta_m))}{8(\mu_{\rm nw} x_m^i + \mu_{\rm w}(L-x_m^i))}, 
\end{equation}
where $q_m$ is the interface velocity, and $\mu_{\rm nw}$ and $\mu_{\rm w}$ are non-wetting and wetting fluid viscosities, respectively. The above expression for interface velocity is equivalent to the rate of change of the interface location $x^i_m$ in tube $m$, which can be defined as ${dx_m^i}/{dt}$. From equation (\ref{eqbundle3}), one can then determine the required time to reach a specified interface position, and subsequently the time to drain each individual tube in a multi-step outflow experiment. 
\subsection{Pore-scale time-dependent wettability model}\label{sec:3}
In this study, the WA agent is defined as either the non-wetting fluid itself or some reactive component therein that is transferred to the wetting fluid across a fluid interface. We consider an alteration process that continues until the ultimate wetting state is reached. The alteration is permanent, but can also be halted at some intermediate wettability state if the WA agent is removed from the system before the final wetting state is reached. If the agent is reintroduced at some later point, alteration continues until the final state.

To this end, we introduce a general functional form of pore-scale WA mechanism,
\begin{equation}\label{contactGnera}
\theta_m(\cdot) := \theta_{m, \rm i} +  \varphi(\cdot){\rm \Delta} {\rm \Theta}, 
\end{equation}
where ${\rm\Delta  \Theta} = \theta_{m,\rm f}-\theta_{m,\rm i}$, $\theta_{m,\rm f}$ and $\theta_{m,\rm i}$ are the ultimate  and initial contact angles respectively.  Note that  $\varphi$ can be any function form that is properly designed to capture the WA processes over a period of time. As discussed in the introduction, the process of contact angle change at the pore-scale involves surface chemistry acting at micro and nano-scale. The mechanistic model in Equation (\ref{contactGnera}) is not intended to capture the full complexity at the pore-scale, but is simply a convenient mathematical form that gives smooth change in contact angle over time between an initial and final contact angle. Detailed laboratory work would needed to further justify the use of this model, which is beyond the scope of this paper.

Though it is well known that  WA  impacts the reservoir system, theoretical investigation and laboratory measurement on time-dependent WA is very limited. However, coupling the instantaneous WA processes to the Langmuir adsorption  model have been successfully implemented in EOR simulation technology \cite{Andersen15, Delshad09, Yu2008, RefLashgari}.  In this paper, the parameter $\varphi(\cdot)$ in equation (\ref{contactGnera}) is assumed to evolve according to Langmuir-type adsorption as a test case.  Thus, $\varphi$ can be defined as follows:
\begin{equation}\label{eq:labda}
\varphi := \frac{\chi_m}{ C +  \chi_m}
\end{equation}
%
where $C$ is non-dimensional  parameter that controls the speed and extent of alteration from  water-wet to intermediate-wet or hydrophobic system. The non-linear variable, $\chi_m$ is a measure of the local exposure time of each individual tube $m$, defined as 
\begin{equation}\label{eqhistflux2}
\chi_m(t,S_{nw_m}) := \frac{1}{T}\int_0^{t}  S_{nw_m}d\tau,
\end{equation}
where $T$ is a pre-specified characteristic time. In this paper, the total time to complete the possible drainage-imbibition cycles from the initial to final wetting state is used as a characteristic time.

For convenience, we may also define $\overline{\chi}$ as a measure of exposure time across the entire bundle (i.e. REV),
\begin{equation}\label{eqhistflux3}
\overline{\chi}(t,S_{nw}) := \frac{1}{T}\int_0^{t}  S_{nw}d\tau.
\end{equation}

Given the above model in Equation (\ref{eq:labda}), we may consider two WA approaches, namely the \textit{uniform} and \textit{non-uniform} alteration models. 

In the case of uniform WA, pores are assumed to be altered identically in time due to transfer of the WA agent by dissolution from neighboring pores. In this case, the local exposure time is equivalent to the average exposure time, such that $\chi_m=\overline{\chi}$ in Equation (\ref{eq:labda}). Alternatively, we may restrict WA to \emph{only} those pores that have been flooded by the non-wetting fluid. In this case, we must keep track of the local exposure time of each tube within the bundle to evaluate Equation (\ref{eq:labda}). This process gives rise to a \emph{non-uniform} or mixed alteration of the pores until such time that all pores have reached the final wetting state. 


\subsection{Simulation approach} 
\label{sub:solution_method}

\begin{figure}[t!]
\centering
\includegraphics[width = 4.5in, height = 4.45in]{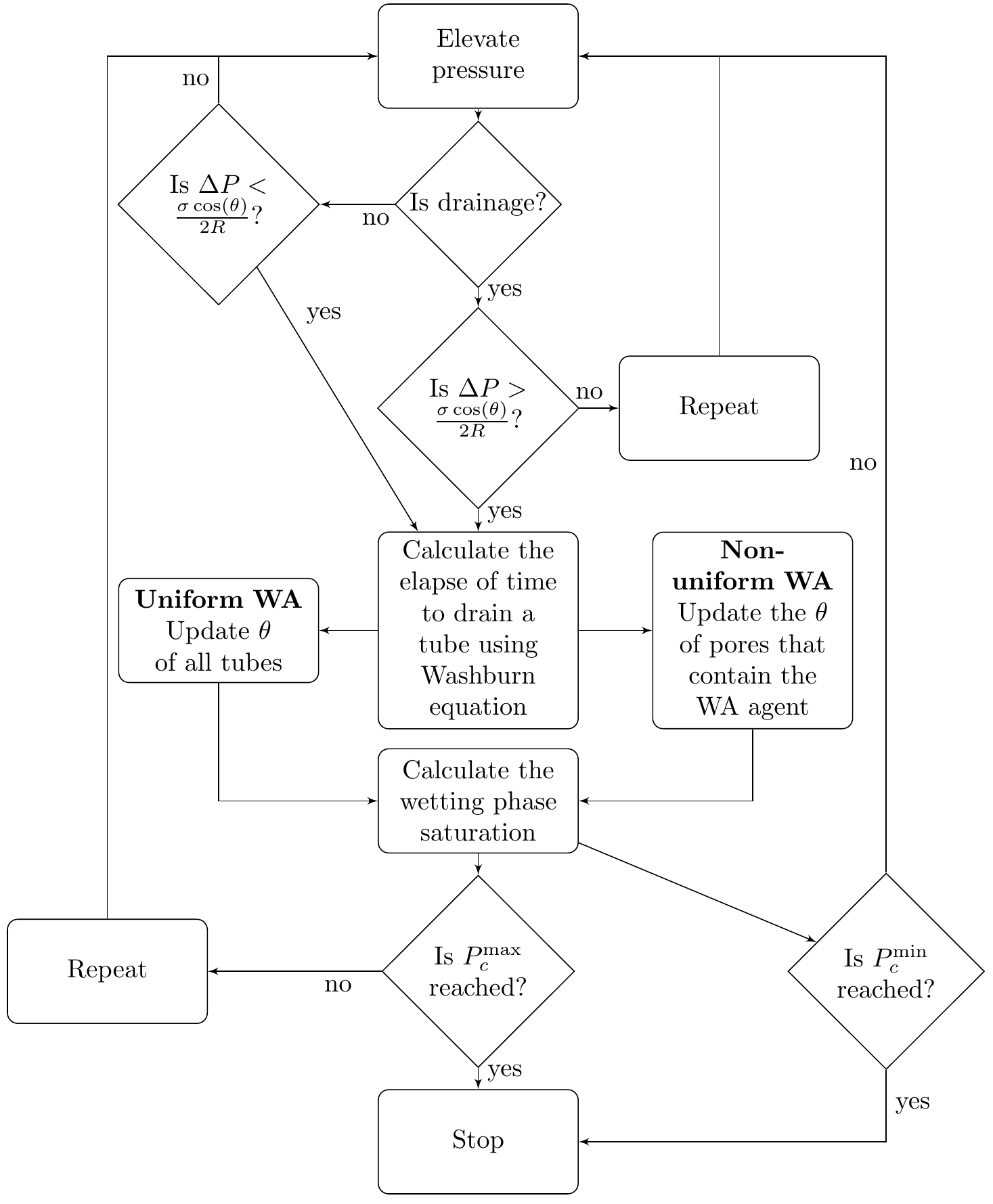}
\caption{Algorithm for wettability alteration and capillary pressure simulation following   drainage-imbibition displacements in a pore-scale model represented by a bundle of tubes.}\label{flow2}
\end{figure}

The uniform and non-uniform approaches for contact angle evolution above are coupled into the BoT model following the algorithm pictured in Figure \ref{flow2}.  
For a drainage cycle, the pressure drop is increased incrementally. Upon each increase, the respective tubes are drained according to the entry pressure inequality.  The Washburn model, Equation (\ref{eqbundle3}), is used to calculate the elapsed time to drain each tube. Afterwards the contact angle (and thus $P_c^{\rm mic}$) is updated according to Equation (\ref{contactGnera}). For the uniform model, the contact angle change is applied uniformly, while in the non-uniform model, the change is applied according to local exposure time. The average saturation is recorded and the cycle is repeated until all tubes have drained. 

During imbibition, the process is reversed. For the non-uniform model, the contact angle continues to change locally until the non-wetting phase is displaced by the wetting phase. For the uniform model, the alteration occurs continuously throughout imbibition according to the average exposure time $\overline{\chi}$.

Repeated drainage and imbibition cycles are run according to this algorithm until the final wetting state is reached. At the completion of the numerical experiment, we obtain a set of \pcs ``data points'' that can be plotted in the usual way. 

\subsection{Correlation approach} 
\label{sub:correlation_approach}
Once the capillary pressure and associated dynamics have been simulated, the resulting curves used to quantify the dynamic coefficient in the interpolation function in Equation (\ref{eq:dynPc_interp}). The goal is develop a correlated model that involves only a single parameter, and this parameter should have a clear relationship to the pore-scale WA process such that changes in the pore-scale WA model parameter $C$ can be directly accounted for in the dynamic coefficient.


We note that for the interpolation model, the dynamics are contained in the coefficient $\omega$ that interpolates between two static capillary pressure curves. Therefore before correlating this model, we determine the functional form of the static wetting states using a simple power-law model. 


\section{Results}\label{sec:4}
In this section, we present the results of simulated capillary pressure experiments using pore-scale models represented by a bundle of tubes following the approach described above. First, the static capillary pressure is simulated for both the initial and final wetting states. Then, simulations are performed on the model subject to WA. Two cases are performed: uniform and non-uniform alteration. The correlated \pcs models for the static cases are presented, followed by the correlated models for the dynamic capillary pressure curves.

\subsection{Bundle of tubes model set-up} 
\label{sub:simulation_set_up}

The pore scale is described by a BoT model (see Section \ref{sec:11}). Each tube in the BoT is assigned a different radius R, with the radii drawn from a truncated two-parameter Weibull distribution \cite{RefHelland}
\begin{equation}
f(R) = \frac{\Big[\frac{R-R_{\rm min}}{R_{\rm av}}\Big]^{\eta-1}\frac{\eta}{R_{\rm av}}{\rm exp}\Big(-\Big[\frac{R-R_{\rm min}}{R_{\rm av}}\Big]^\eta\Big)}{1-{\rm exp}\Big(-\Big[\frac{R_{\rm max} - R_{\rm min}}{R_{\rm av}}\Big]^\eta\Big)}
\end{equation}
where $R_{\rm max}$, $R_{\rm min}$, and $R_{\rm av}$ are the pore radii of the largest, smallest, and average pore sizes, respectively, and $\eta$ is a dimensionless parameter.  The rock parameters and fluid properties are listed in Table \ref{BtubeM}.
\begin{table}[!ht]
\centering
\begin{tabular}{l l l l l l}
\hline
parameters           & values & unit & parameters  & values & unit\\
\hline
$\sigma_{ow}$        & 0.0072   & {\rm N/m}      & no. radii               & 500    & [-]\\
$R_{\rm min}$        & 10        & {\rm  $\mu$m}       & $R_{\rm max}$     &  40   & {\rm  $\mu$m}\\
$\theta_f$           & 80        &    degree        & $\theta_i$ & 0.0    &    degree\\
$\mu_{\rm w}$        & 0.0015    & {\rm Pa.s}     & $\mu_{\rm nw}$         & 0.0015 & {\rm Pa.s}\\
$r_{\rm av}$         & 23 & {\rm  $\mu$m}&L             & 0.001    & {\rm  m} \\
$\eta$               & 1.5 & [-]         &   & &\\
\hline
\end{tabular}
\caption{Parameters used to simulate quasi-static fluid displacement in BoT. }\label{BtubeM}
\end{table}
These parameters are coupled to the BoT model to simulate capillary pressure curves. 


%

\subsection{Static capillary pressure for end wetting states}\label{staticCorelation}

A starting point for the dynamic capillary-pressure models presented in Section 2 (Equations (\ref{eq:dynPc}) and (\ref{eq:dynPc_interp})) is characterizing the capillary pressure curves for the end wetting states. Given the same tube geometry and fluid pairing described above, the capillary pressure functions are simulated under static conditions for both the initial and final wetting states using the BoT model. 

Only a single drainage experiment is needed in the static case to fully characterize the capillary pressure curve. This is due to the simplicity of the BoT model and lack of residual saturation in cylindrical tubes. Therefore, hysteresis is not possible if the contact angles in the tubes (and other parameters) are held constant.

\begin{figure}[h!]
\centering
\includegraphics[scale=0.6]{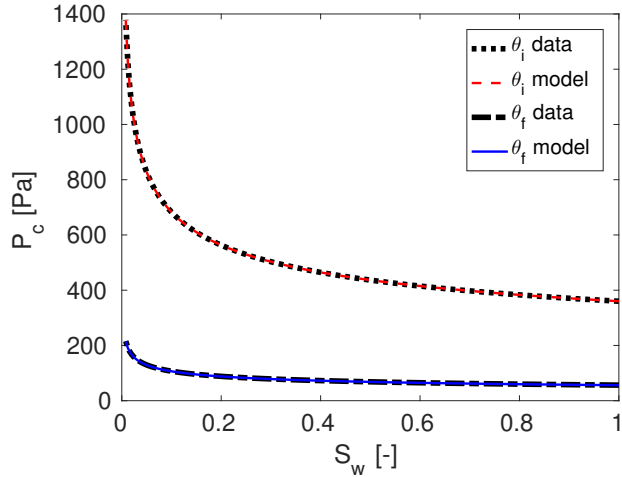}
\caption{Simulated \pcs data for the initial and final wetting states compared to a Brooks-Corey model with calibrated parameters given in Table \ref{tab:paralist}.}\label{fig:corre1}
\end{figure}

The simulated static curves
can be correlated with the Brooks-Corey model (\ref{eq:bc1}). The resulting correlations can be found in Figure~\ref{fig:corre1}, while fitted parameters for the Brooks-Corey model can be found in  Table \ref{tab:paralist}.  
\begin{table}[h!]
\centering
\begin{tabular}{p{1.281cm} p{1.28cm} p{1.28cm}   p{1.28cm}  p{1.28cm} p{1.28cm}}
 \hline
 \multicolumn{2}{l}{\underline{Initial wetting state}} &  & \underline{Final wetting state} \\[0.1in]
param.    & value     & unit   & param.    & value   & unit   \\\hline
  $c_w$    &  360        &  [Pa]   &  $c_w$       &  56      &  [Pa] \\ 
 $a_w$     &  0.2778     &   [-] & $a_w$    & 0.2778     &  [-] \\ 
$ \mathrm{R}^2$  &  1  &    -    & $\mathrm{R}^2$     &  1  &   - \\
 \hline
 \hline
\end{tabular}
\caption{Estimated correlation parameter values for initial and final wetting state capillary pressure curves}\label{tab:paralist}
\end{table}
Figure \ref{fig:corre1} compares the Brooks-Corey formula (\ref{eq:bc1}) and the capillary pressure curves associated with static contact angles (initial and final wetting states).

The Brooks-Corey correlation gives an excellent match to the simulated \pcs data under static conditions.
We observe that the pore-size distribution index $a_w$ for the initial and final wetting states are the same for both end wetting states, which is expected since the same distribution of tube radii is used in both cases. While, on the other hand, the coefficient $c_w$ decreases by a factor of 0.85 from the initial to final wetting state, corresponding to a decrease in macroscale capillary entry pressure. Leverett-J scaling theory \cite{RefXu} predicts that entry pressure scales by $\cos\theta$, which agrees nicely with the reduction in $\cos\theta$ by a factor of 0.83 for a contact angle change from 0 to 80 degrees.

We reiterate that for the static case where no wettability alteration occurs, the Brooks-Corey model describes \emph{both} drainage and imbibition for the bundle of tubes. 

\subsection{Dynamic capillary pressure for uniform wettability alteration}\label{uniformWA}

\begin{figure}[t!]
\centering
\includegraphics[scale = 0.65]{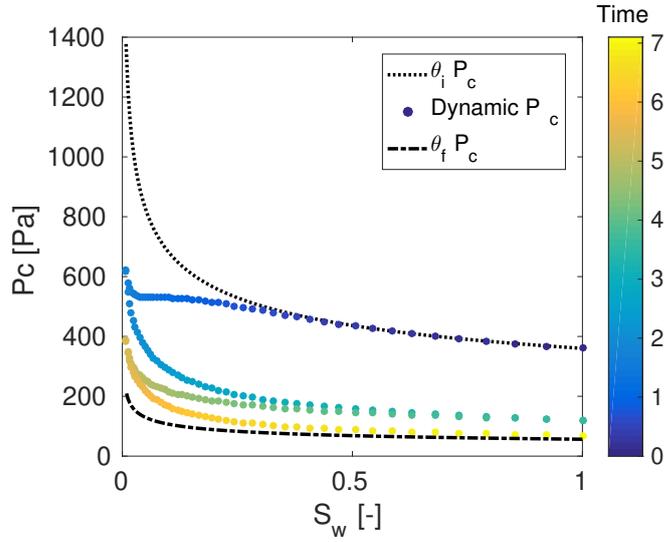}
\caption{Simulation results for WA-induced dynamics of capillary pressure as a function of wetting saturation given uniform wettability alteration at the pore scale. The color of each data point indicates the time elapsed in months. The static \pcs curves for the end wetting states, $\theta_i$ and $\theta_f$, are plotted as a reference (the black broken lines).}\label{fig4:2}  
\end{figure}

Here, we present results for the first dynamic case (Figure \ref{fig4:2}) in which wettability is altered uniformly throughout the bundle by considering the pore-scale parameter $C=0.005$. Two dynamic drainage-imbibition cycles are simulated using the quasi-static BoT approach over a period of approximately 7 months.  
During this time, the capillary pressure steadily decreases from the initial to the final wetting state with each subsequent cycle. Having reached the final wetting state, any additional drainage-imbibition cycle would follow along the static curve for the final wetting state. We observe that wettability-induced dynamics also introduces an apparent hysteresis in the \pcs data. We recall that a simple BoT model cannot exhibit hysteresis under static wettability conditions. 

\begin{figure}[t!]
\centering
\includegraphics[width=0.65\textwidth]{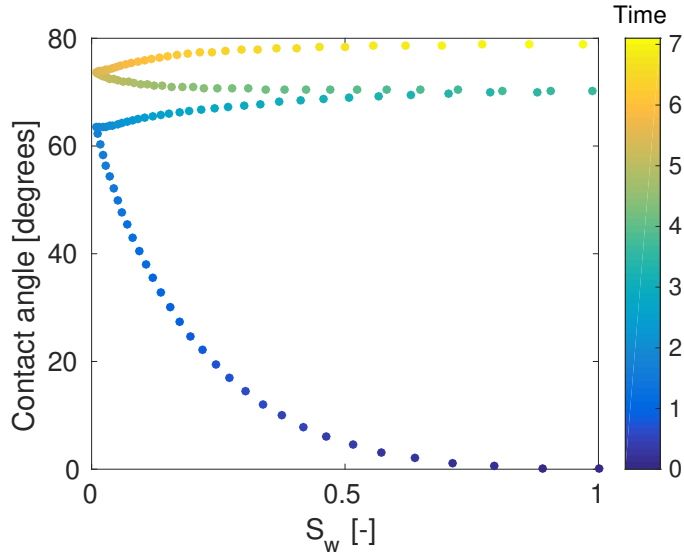}
\caption{Uniform contact angle $\theta$ change across the bundle as a function of drainage-imbibition fluid history paths. The color of each data point indicates the time elapsed in months. }\label{fig4:1} 
\end{figure} 

The dynamics in capillarity are coupled to a continuous change in contact angle between the two wetting states, which is shown in Figure~\ref{fig4:1} for the same two cycles.  
We observe that $\theta$ undergoes a rapid increase by the end of the first drainage, followed by a slower evolution towards the final wetting state. Since there is no spatial variability in wettability at the pore scale in the uniform case, smaller tubes are altered at the same rate as larger tubes. This is evident in the associated \pcs data where we observe substantial alteration of the first drainage path at low wetting saturation. The alteration from the initial wetting condition is fast, leading to non-trivial and almost non-monotonic dynamics in capillarity at early time. 

\begin{figure}[t!]
\centering
\includegraphics[scale=.45]{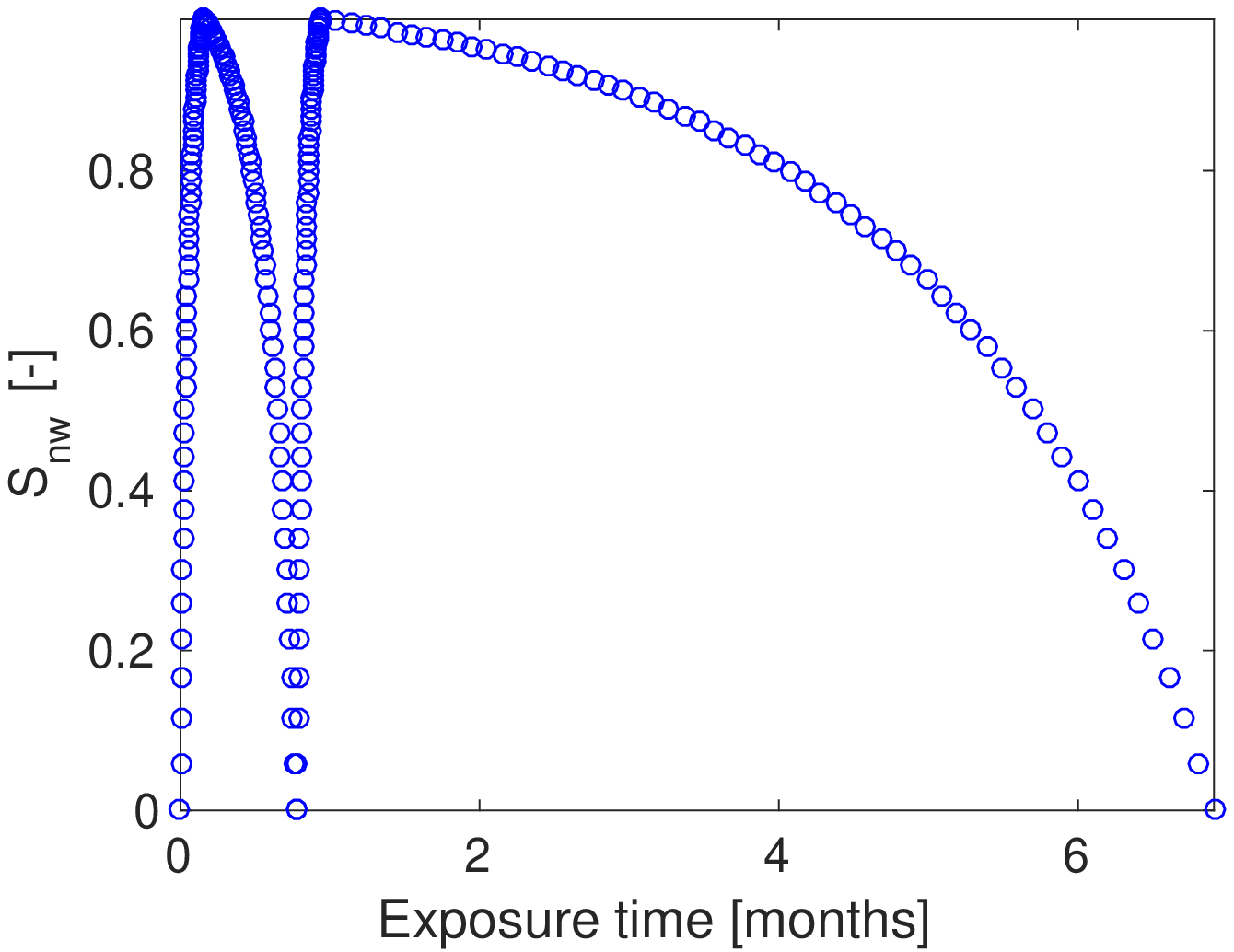}
\includegraphics[scale=0.45]{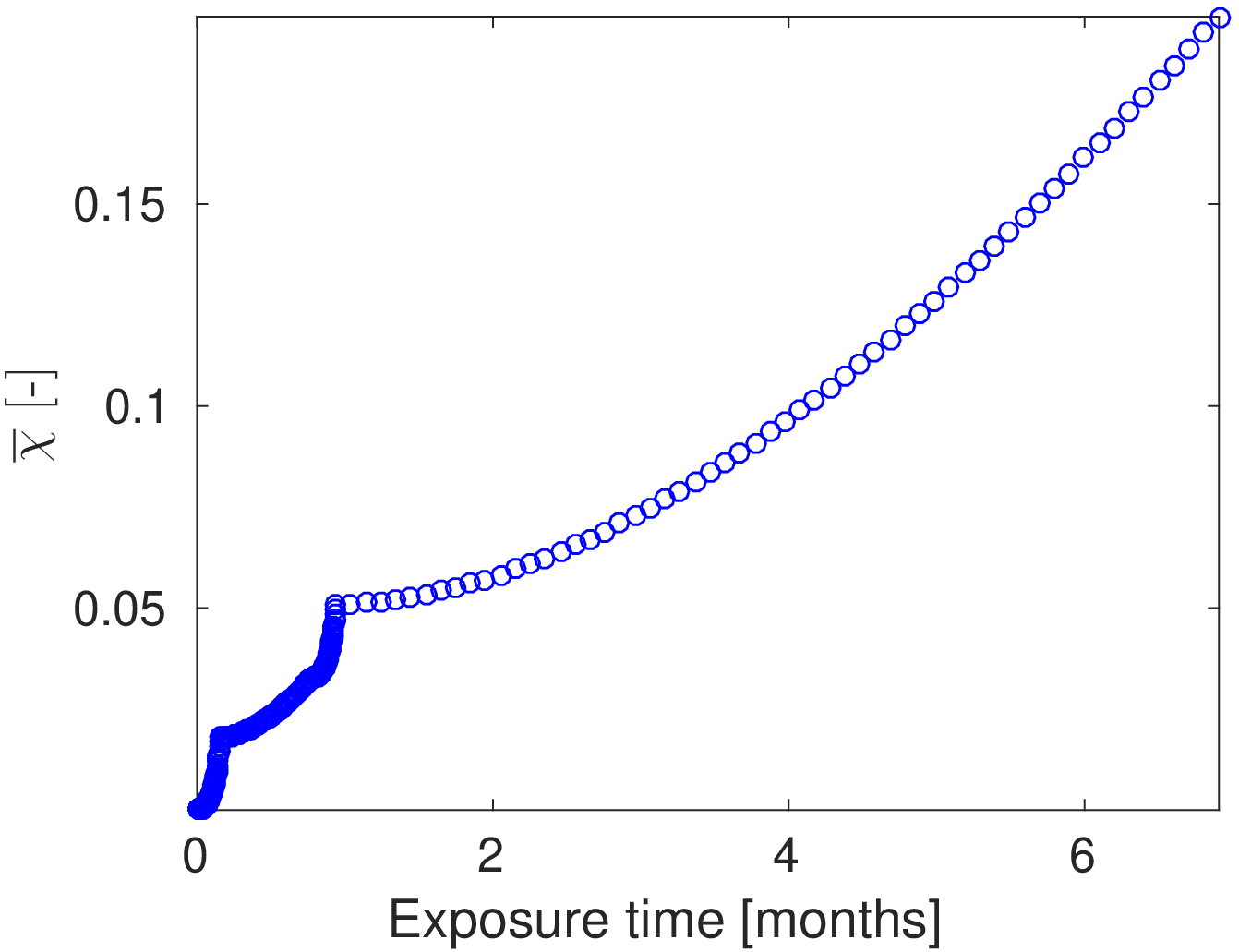}
\caption{Saturation history for the simulated \pcs data over two drainage-imbibition cycles as shown in Figure \ref{fig4:2} for the uniform WA case: (a) average saturation history of the non-wetting fluid with respect to time in months, and (b) the integrated exposure time $\overline\chi$ over the same period.}\label{fluidHis}
\end{figure}

The saturation history is an important factor, defining the length of exposure in  individual tubes. Figure \ref{fluidHis}a shows that saturation history is driven by the time required to displace fluid through each tube in sequence, i.e. from largest to smallest during drainage and vice versa during imbibition. From integration of this curve, we obtain the exposure time $\overline\chi$ (Figure \ref{fluidHis}b) which controls the contact angle change shown previously in Figure~\ref{fig4:1}. We recall that for the uniform model, contact angle change at the pore-scale (Equation (\ref{eq:labda})) is driven by the average saturation not the local saturation. 

Figure \ref{fluidHis}a also indicates the time taken for each drainage-imbibition cycle, with the first cycle takes approximately 1 month to complete. Since the displacement process is speeding up with increasing contact angle, up to 12 cycles would be required in this manner before the end wetting state is reached. In the interest of presentation, we have controlled the pressure drop during the final imbibition experiment so that exposure time was lengthened in order to reach the final wetting state by the end of the second cycle.

\begin{figure}[t!]
\centering
\includegraphics[width=0.48\textwidth]{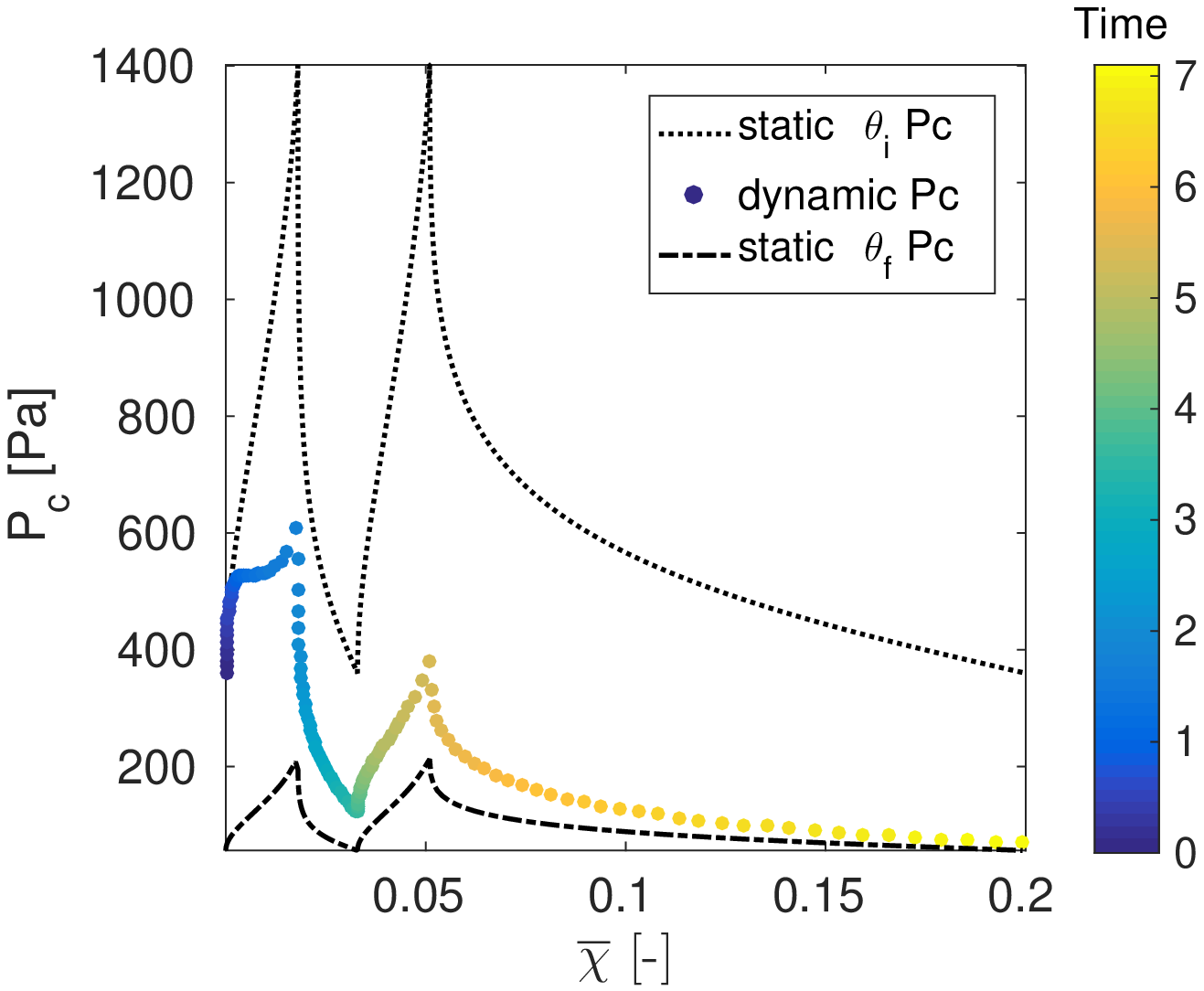}
\includegraphics[width=0.48\textwidth]{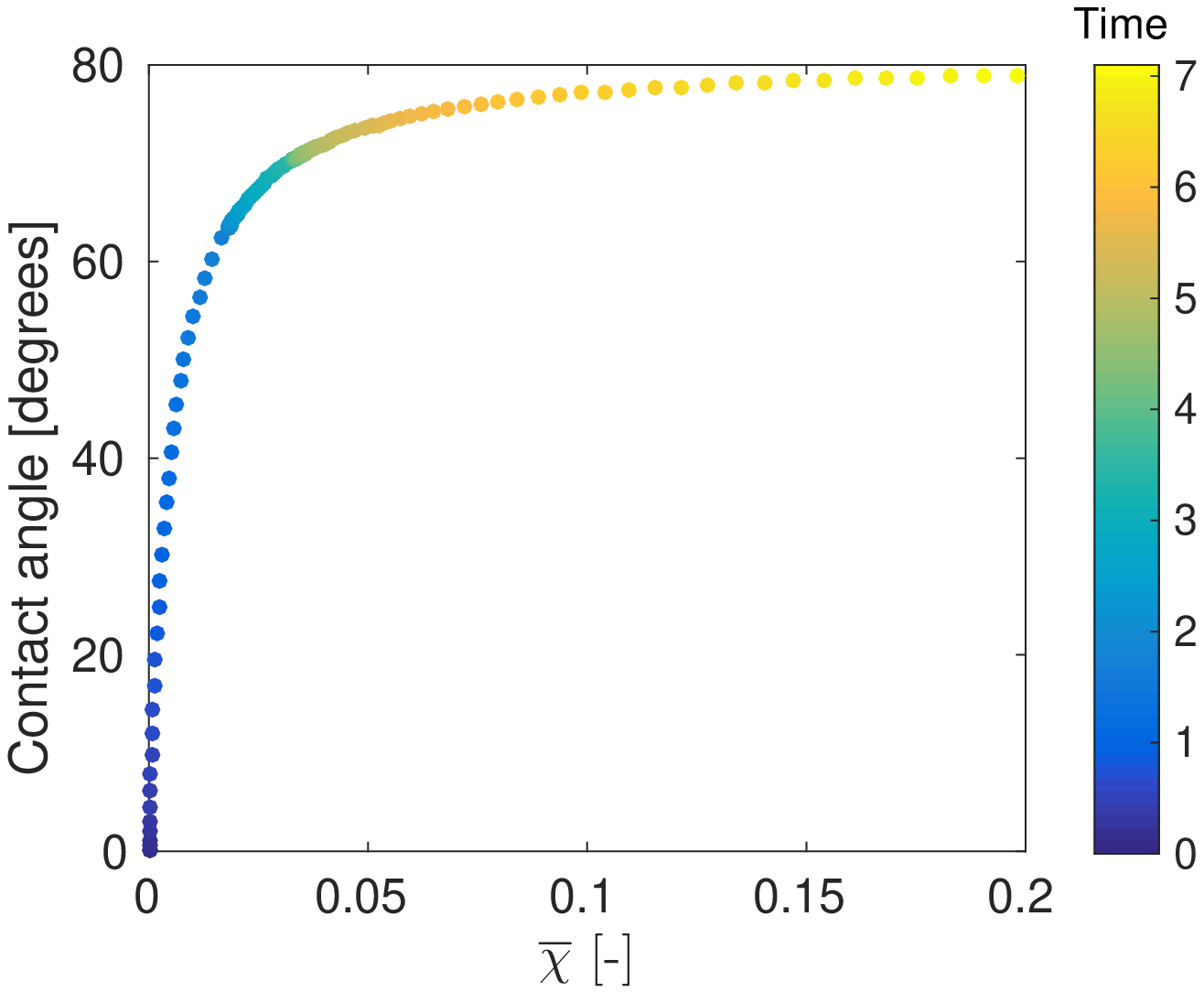}
\caption{The capillary pressure data (a) and contact angle (b) as a function of  $\overline{\chi}$ for the uniform WA case. The color of each data point indicates the time elapsed in months.}\label{fig4:1b} 
\end{figure}

We have observed above that capillary pressure is no longer a unique function of saturation as it is in the static case for this simple geometry. However, if \pcs is transformed to the temporal domain by plotting against $\overline\chi$, we obtain a unique function with respect to exposure time. However, Figure~\ref{fig4:1b}a shows that the capillary pressure relates with the temporal domain $\overline{\chi}$ non-monotonically  i.e., the capillary pressure  increases and decreases with each drainage-imbibition cycle. 
Similarly, the associated contact angle change is plotted along with  exposure time (Figure~\ref{fig4:1b}b), which is exactly the function given in Equation~(\ref{eq:labda}) at the pore level. The value of $\theta$ is smoothly increasing in time according to the pore-scale model.

\subsubsection{Dynamic capillary pressure model development} 
\label{sub:correlated_functions_for_dynamic_pc_s_uniform_case}
\begin{figure}[t!]
\centering
\includegraphics[width=0.48\textwidth]{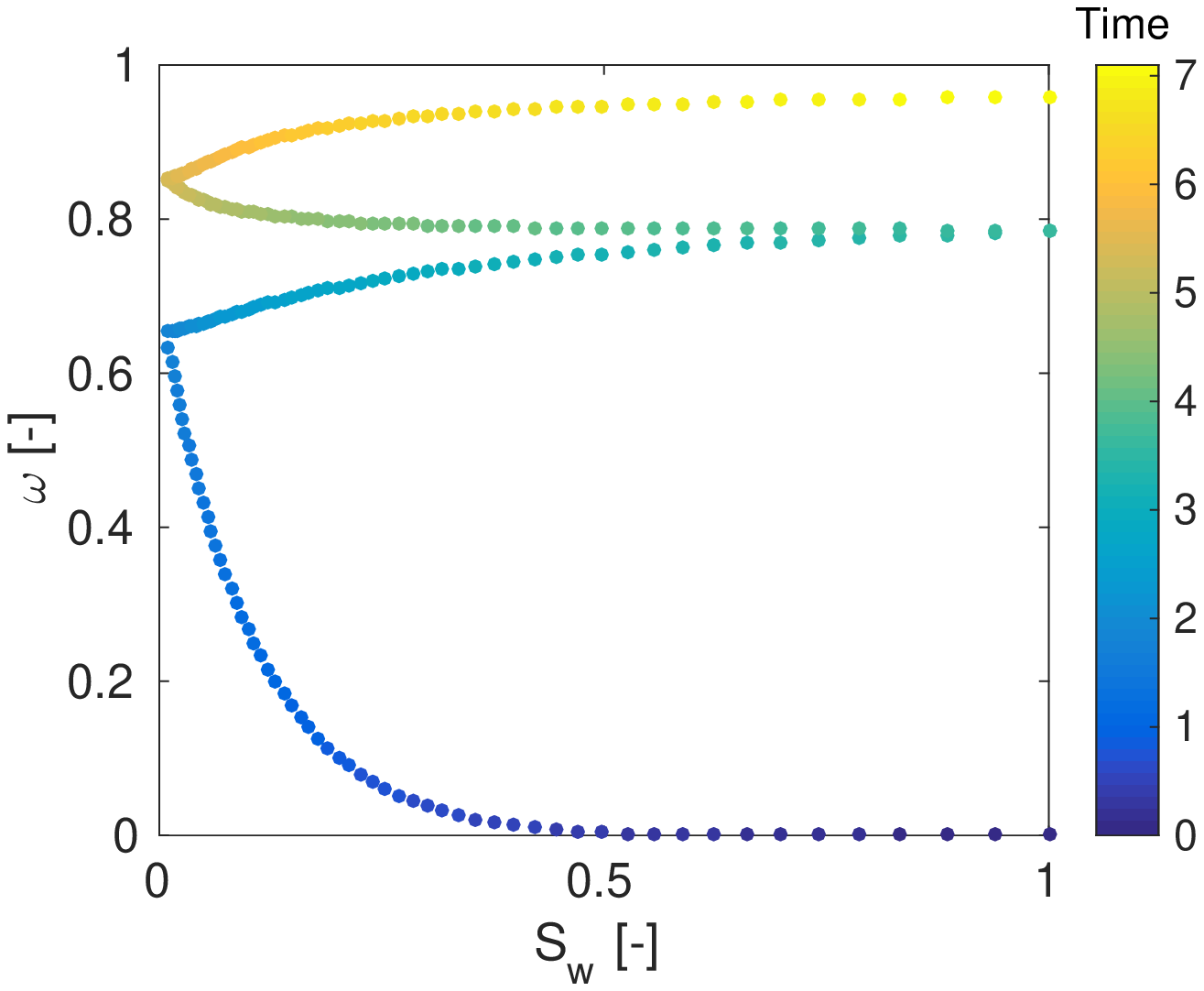}
\includegraphics[width=0.48\textwidth]{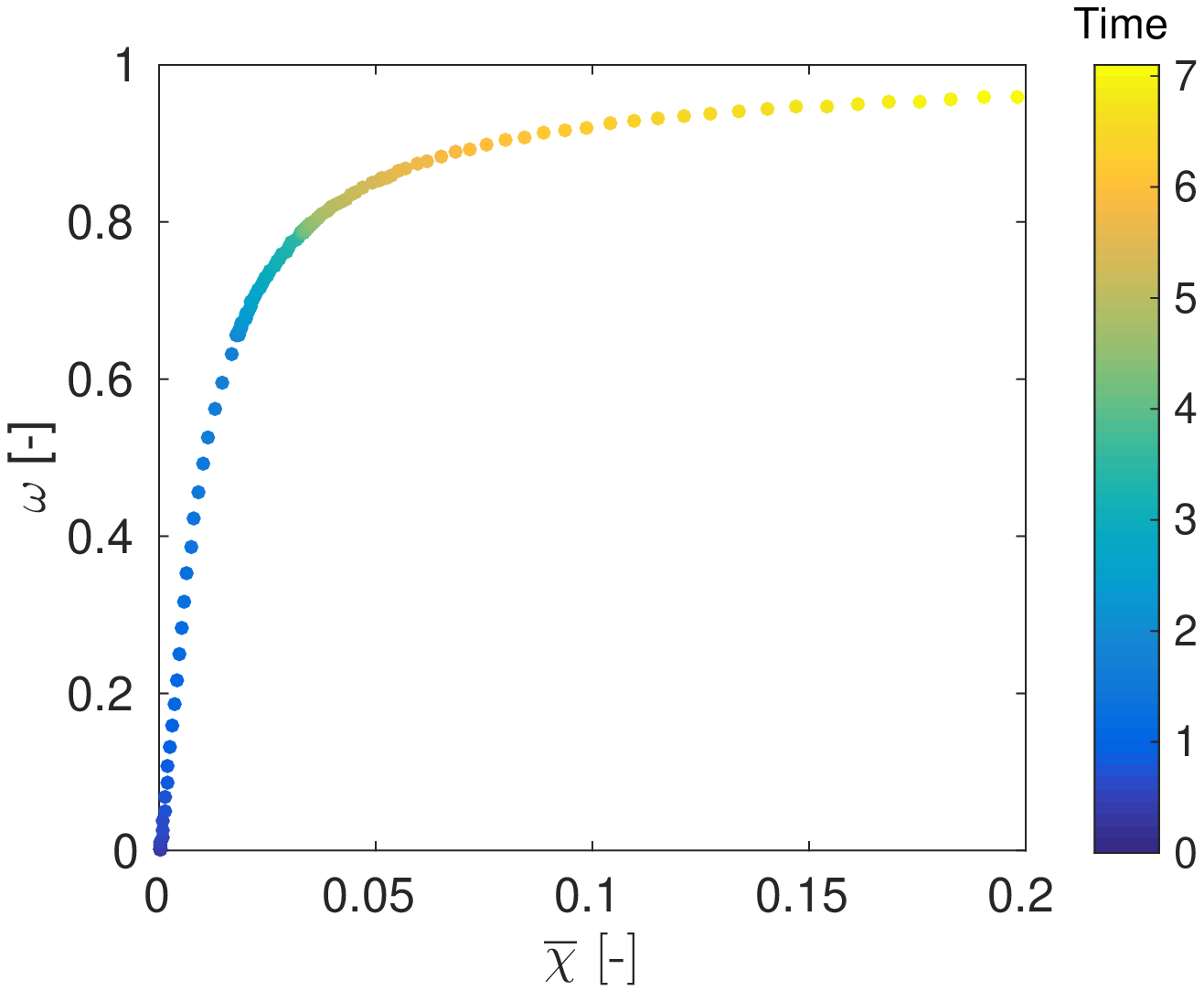}
\caption{Plot of the dynamic coefficient $\omega$ against wetting phase saturation (a) and  $\overline{\chi}$ (b) for the uniform wettability alteration case. The data points are color-coded with exposure time in months.}\label{fig4:11}
\end{figure}

Following the approach discussed in Section \ref{sec:approach},  we  applied Equation (\ref{eq:dynPc2}) to determine the dynamic coefficient $\omega$. As shown in Figure \ref{fig4:11}a, we observe that $\omega$ is a non-unique function of wetting-phase saturation but with values that are continuously increasing as the dynamic capillary pressure approaches the final wetting state. This non-uniqueness behavior makes challenging to propose a functional form for $\omega$-$S_w$ relation. 

On the other hand, Figure \ref{fig4:11}b shows that $\omega$ is smoothly increasing, uniquely related to $\overline{\chi}$, and mimics the functional form of the pore-scale model in Equation (\ref{contactGnera}) (see Figure \ref{fig4:1b}b).
Given this insight, we propose an adsorption-type model to fit the interpolation coefficient $\omega$ to the average exposure time:
\begin{equation}\label{eqnmoduni3d}
\omega = \frac{\overline{\chi}}{\beta_1 + \overline{\chi}},
\end{equation}
where $\beta_1$ is a fitting parameter obtained from the best fit to simulated data in Figure~\ref{fig4:11}b. For this particular case, the calibrated parameter is estimated to be $\beta_1 = 0.01$.

\begin{figure}[t!]
\centering
\includegraphics[scale=.45]{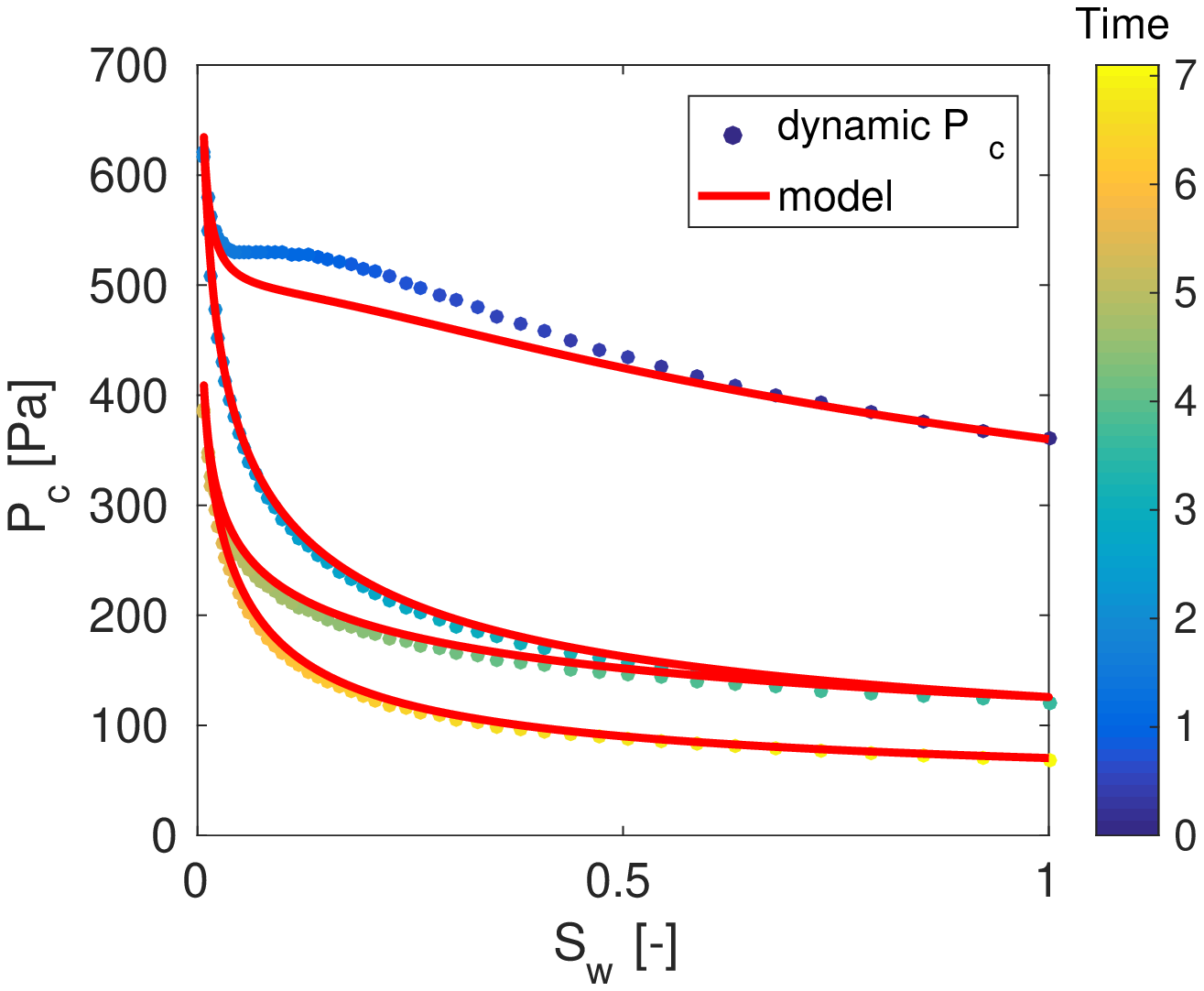}
\includegraphics[scale=0.45]{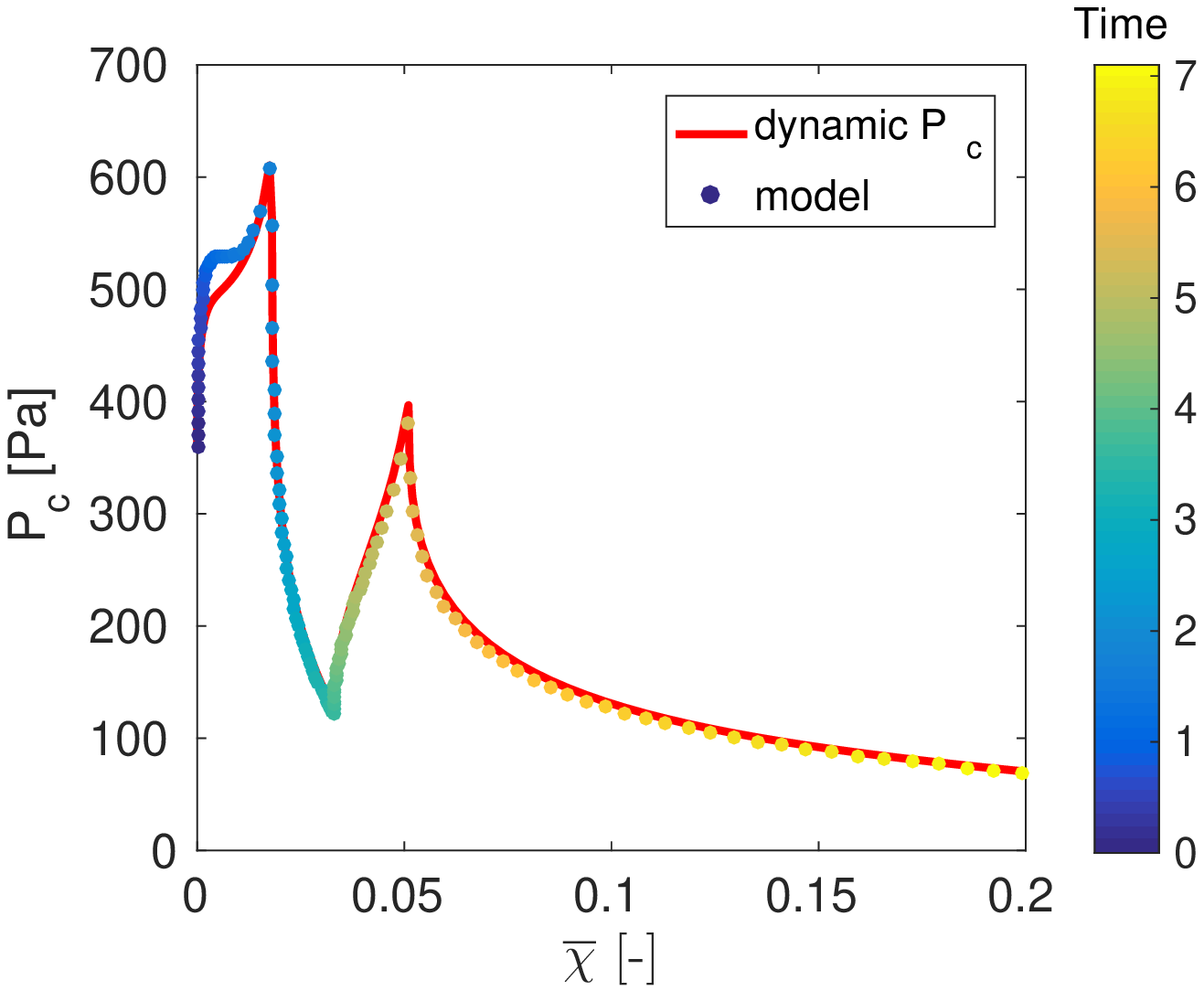}
\caption{Comparison of the dynamic capillary pressure model for uniform wettability alteration with simulated data: as a function of wetting saturation (a) and $\overline{\chi}$ (b).}\label{fig:cor2}
\end{figure}

The proposed model for $\omega$ in Equation~\eqref{eqnmoduni3d} is incorporated into the dynamic capillary pressure model proposed in Equation~\eqref{eq:dynPc_interp} and compared against the simulated dynamic capillary pressure curves. The resulting fit, shown in Figure \ref{fig:cor2} indicates that a one-parameter model for the dynamic coefficient agrees well with the simulated data. The correlation coefficient for this comparison is $R^2 = 0.9921$.

\begin{figure}[h!]
\centering
\includegraphics[scale=0.55]{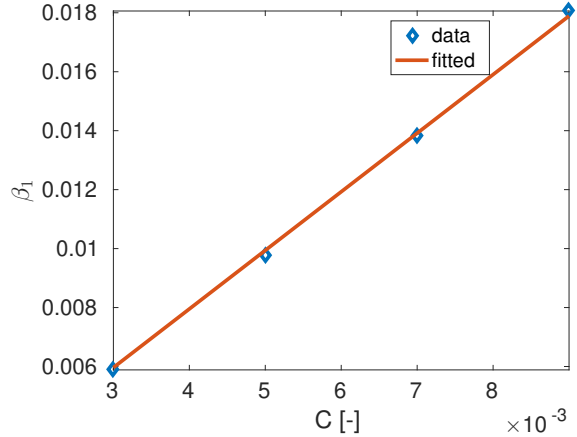}
\caption{The relation between pore-scale wettability model parameter $C$ and the   dynamic coefficient parameter $\beta_1$.}\label{pararela}
\end{figure}

We hypothesize that the value of $\beta_1$ in this model is dependent on the dynamics of wettability alteration at the pore-scale. Therefore, we investigate how the parameter $\beta_1$ evolves with changes in the pore-scale parameter $C$ in Equation (\ref{contactGnera}). We repeat the capillary pressure simulations for different values of the pore-scale parameter $C$ and determine the correlated value of $\beta_1$ in each case. Figure \ref{pararela} shows that the interpolation model parameter is directly proportional to the pore-scale model parameter, with a proportionality constant of 2. Thus, the relationship $\beta_1 = 2C$ can be used to predict the macroscale parameter directly from knowledge of the pore-scale process. 

The final dynamic capillary pressure model can now be obtained by incorporating the above model for $\omega$  with the established relationship for $\beta_1$ into the general form of the interpolation model in Equation~\eqref{eq:dynPc_interp}:
\begin{equation}
    P_c= \frac{\overline{\chi}}{2C + \overline{\chi}} \Big(P_c^{\rm st,f} - P_c^{\rm st,i} \Big)+ P_c^{\rm st,i}.
    \label{eq:dynPc_interp_re}
\end{equation}
In its final form, the dynamic capillary pressure model in Equation~\eqref{eq:dynPc_interp_re} is solely a function of the pore-scale parameter $C$. This parameter must be determined from laboratory experiments for a given sample exposed to a WA agent.

\subsubsection{Dynamic model applicability} 
\label{sub:model_applicability}

We note that the saturation history in Figure~\ref{fluidHis}a used to generate the \pcs curves in Figure~\ref{fig4:2} can be thought of as one arbitrary path of an infinite number of possible paths. If a different path had been chosen, such as a flow reversal at intermediate saturation or a prolonged exposure time at a given saturation, it would result in entirely different capillary pressure dynamics. 

\begin{figure}[t!]
\centering
\includegraphics[scale=0.45]{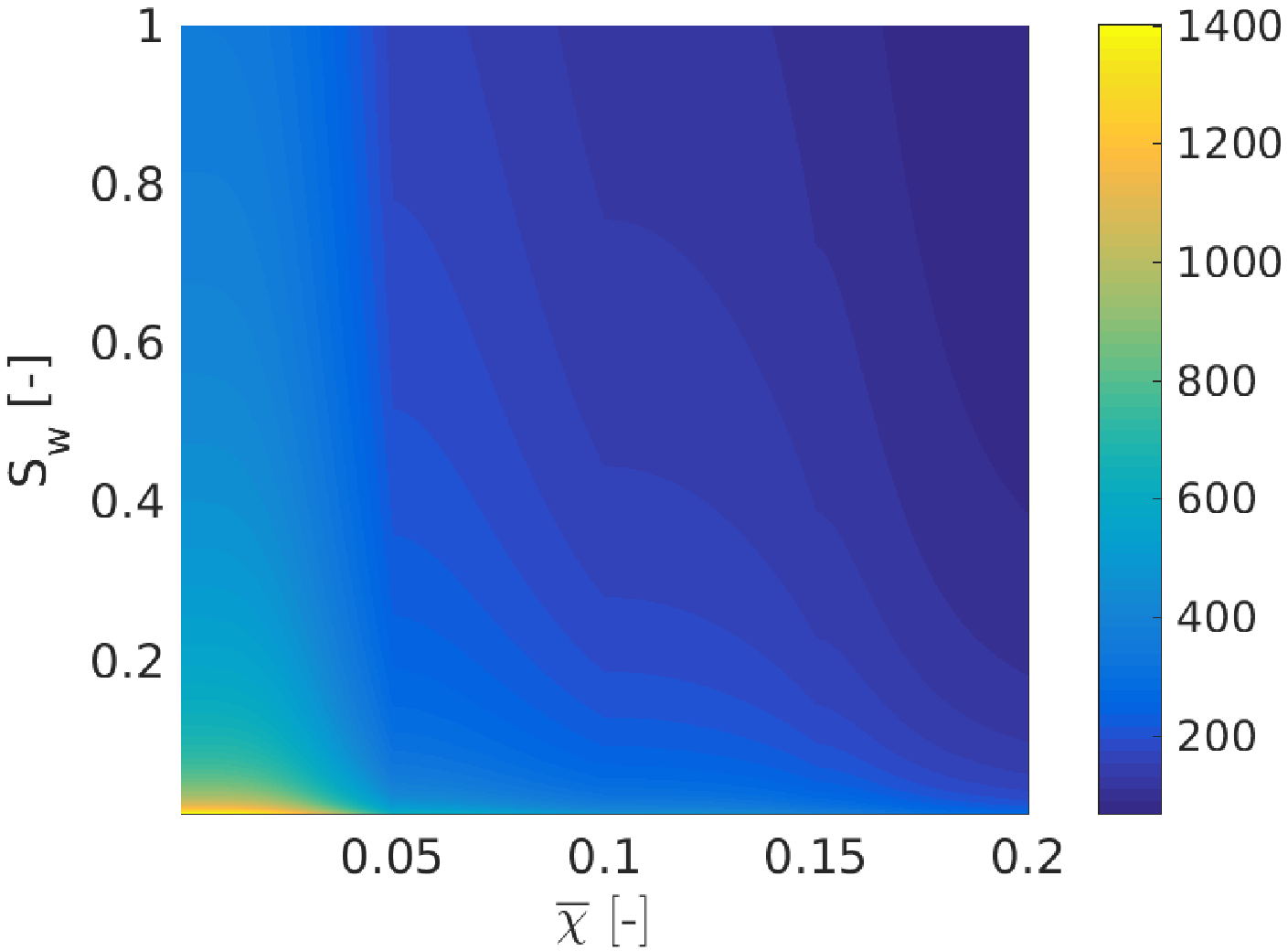}
\includegraphics[scale=0.45]{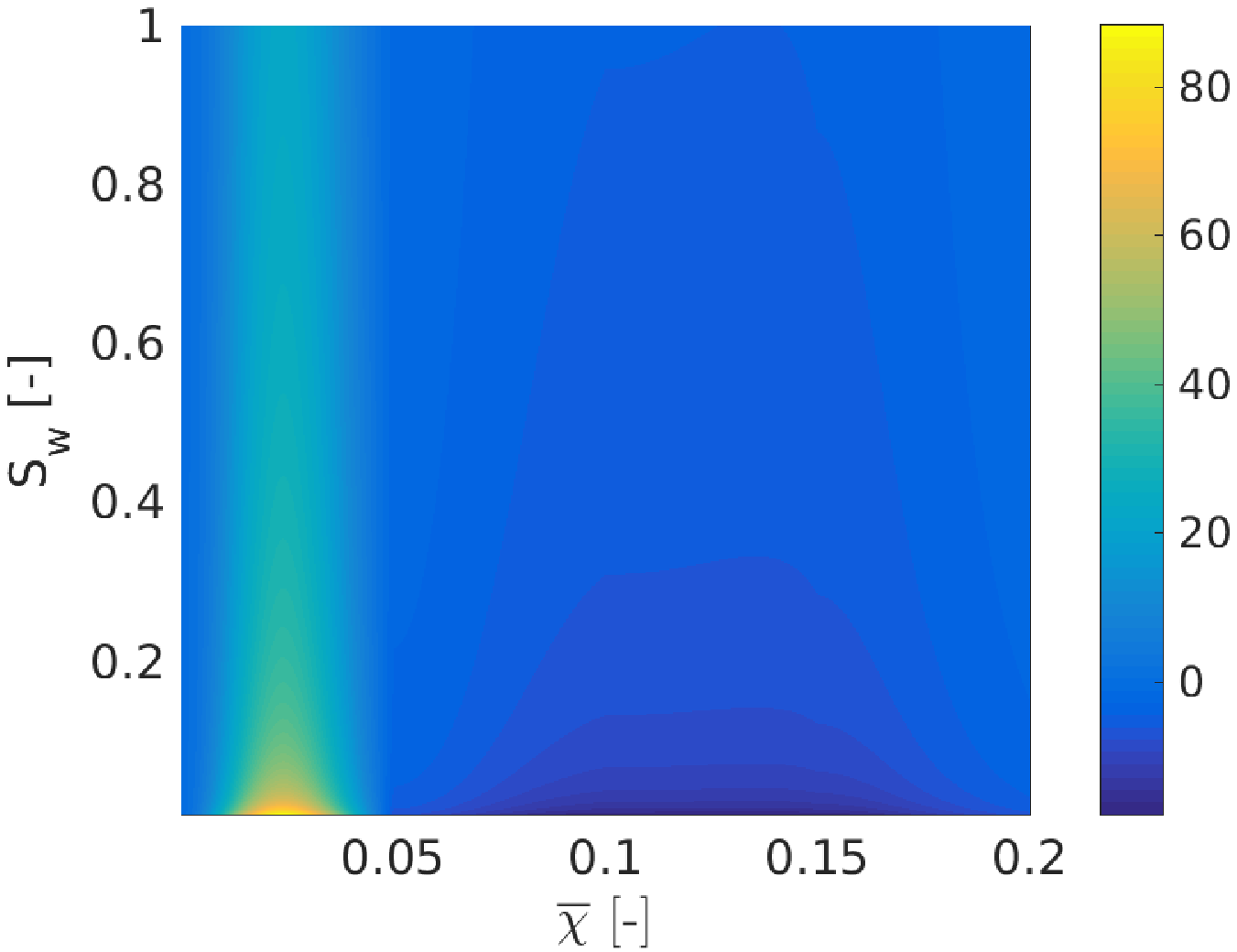}
\caption{(a) Simulated capillary pressure under uniform WA obtained by taking multiple paths in the $S_w\times \overline{\chi}$ domain. (b) Difference between the dynamic capillary pressure model prediction (\ref{eq:dynPc_interp_re}) with the data in (a).}\label{fig:dycapSurf-core}
\end{figure}

We generate many different \pcs curves by taking numerous different paths in the saturation-time domain. The resulting simulated data forms a surface with respect to saturation and exposure time as shown in Figure~\ref{fig:dycapSurf-core}a. 
We then apply the calibrated dynamic model from Equation (\ref{eq:dynPc_interp_re}) to the same saturation-time paths used to generate the $P_c$-$S$-$\overline{\chi}$ surface. The difference between the calibrated model and the simulated data is shown in Figure~\ref{fig:dycapSurf-core}b. A good comparison of the dynamic model to simulated data demonstrates that model calibration to a single saturation-time path is robust enough to be applied to any possible path. 

\subsection{Dynamic capillary pressure for non-uniform wettability alteration}\label{sec4.2}
\begin{figure}[t!]
\centering
\includegraphics[scale=0.65]{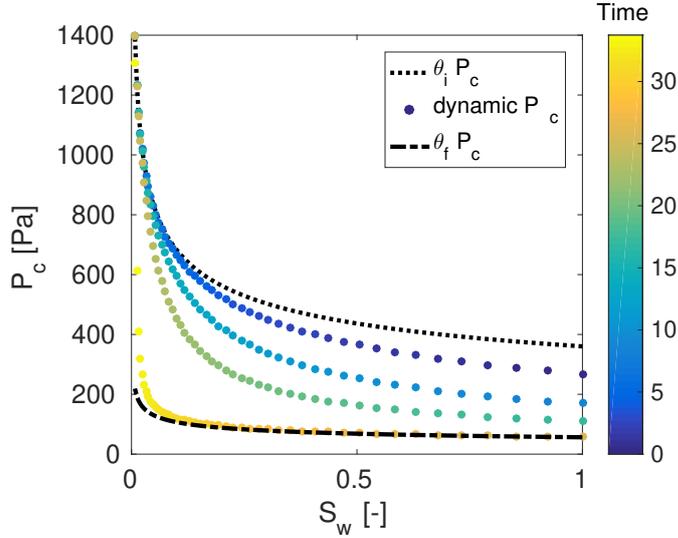}
\caption{Simulated capillary pressure curves for the case of non-uniform WA over four drainage-imibition cycles with respect to wetting phase saturation.  The static \pcs curves for the initial and final wetting states are shown as a reference. The color code shows the \pcs dynamics along  approximatly 33 months of exposure time.}\label{fig4:4s}
\end{figure}

We present the second case where WA evolves   throughout the bundle  non-uniformly at the pore level with the pore-scale parameter $C=1\times10^{-5}$. In this case, we performed four WA induced dynamic drainage-imbibition displacements and are shown in Figure \ref{fig4:4s}.
The first drainage curve follows exactly along the static curve for the initial wetting state, which is expected since the entry pressure to each individual tube is not altered before the tube has drained. Only after the tubes are drained by the WA agent can the contact angle change in each tube. Therefore, it is only the imbibition data that show the corresponding reduction in capillarity. At the end of imbibition, the system becomes quasi-static in the sense of alteration, which means the wettability of a given tube stops being further altered until a new drainage cycle has begun. This implies that each new drainage curve must follow along the previous imbibition curve and thus, imbibition and drainage paths overlap in Figure \ref{fig4:4s}. This effect is in contrast to the uniform WA case, where each drainage and imbibition curves follow different paths. 

\begin{figure}[t!]
\centering 
\includegraphics[scale=.65]{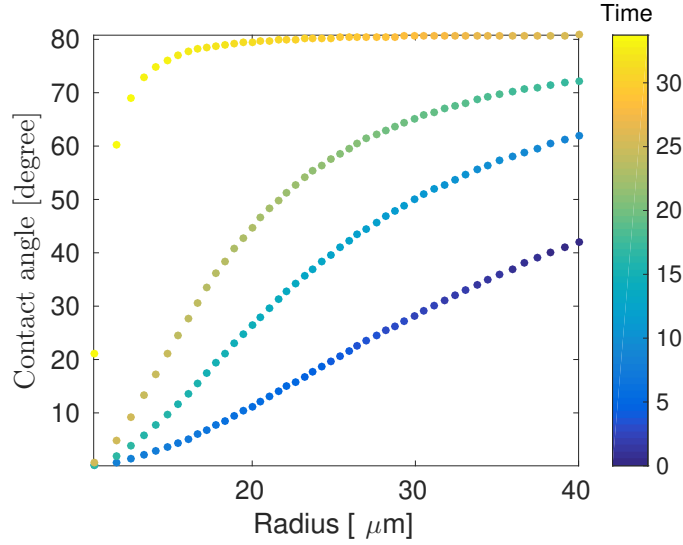}
\caption{Dynamic contact angle evolution as a function of exposure time to the WA agent per each tube. This contact angle distribution was recorded at the end of each drainage-imbibition cycle and the color code shows the contact angle dynamics during 33 months of exposure time.}\label{fig4:4}
\end{figure}

The capillary pressure instability in Figure \ref{fig4:4s} is attributed to the local change of wettability in each tube. 
The larger tubes drain first and are imbibed last, resulting in a long local exposure time compared to the smaller tubes that drain last and imbibe first. Therefore, the initial wetting state persists in the smaller tubes.  Figure \ref{fig4:4} shows the resulting heterogeneous contact angle change in time across the bundle. 

We observe that the impact of fluid history and exposure time on the wetting property of the bundle and the associated capillarity is distinct from the uniform WA case. Here, the dynamic capillary pressure continues to rise to the initial wetting state capillary pressure curve for low water saturation. This is due to the reason that the smaller tubes were exposed to the WA agent for shorter time than the larger tubes.  This situation persists until  the bundle approaches a uniform wetting condition  corresponding to the final wetting state.

\begin{figure}[t!]
\centering
\includegraphics[scale=0.45]{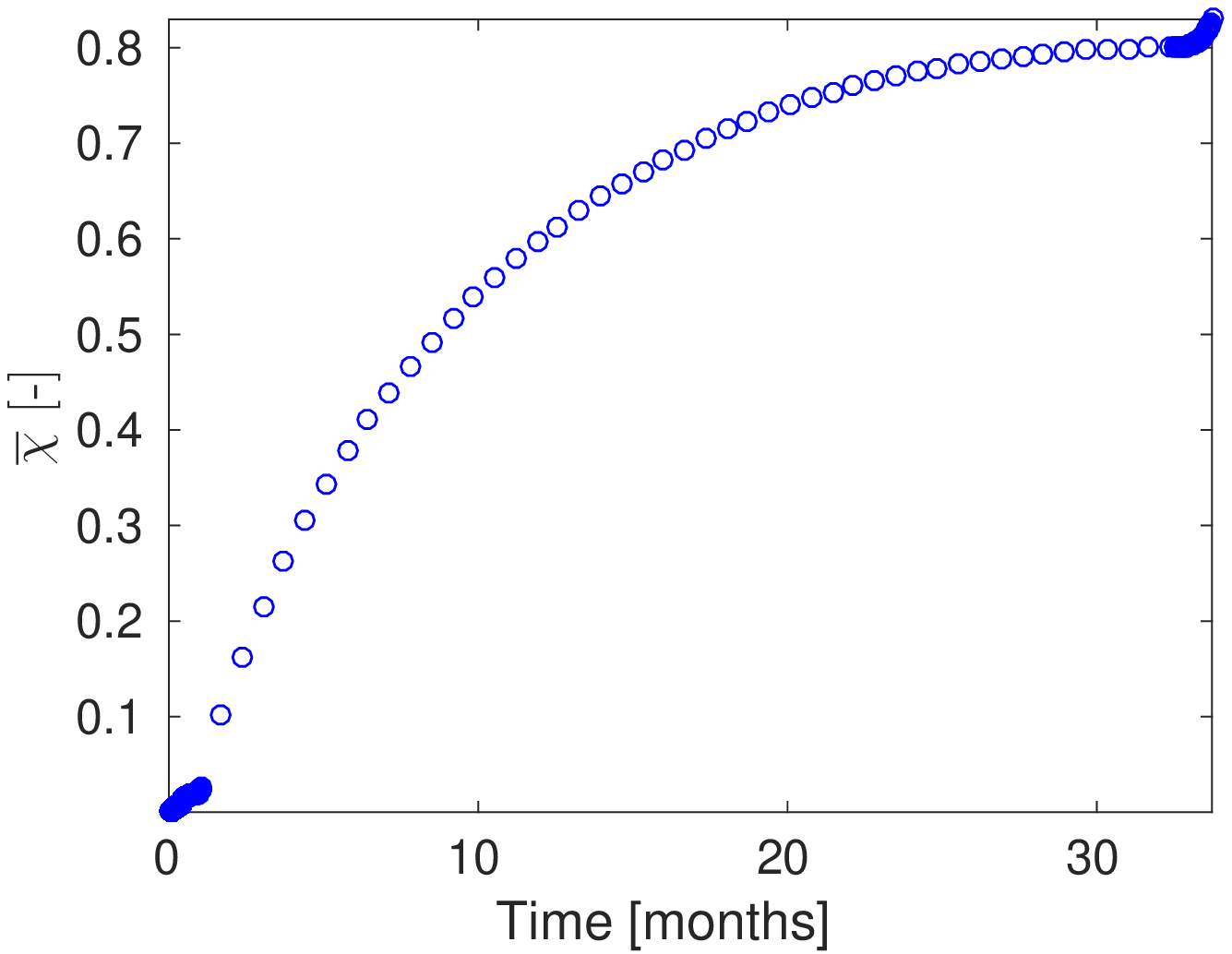}
\includegraphics[scale=0.45]{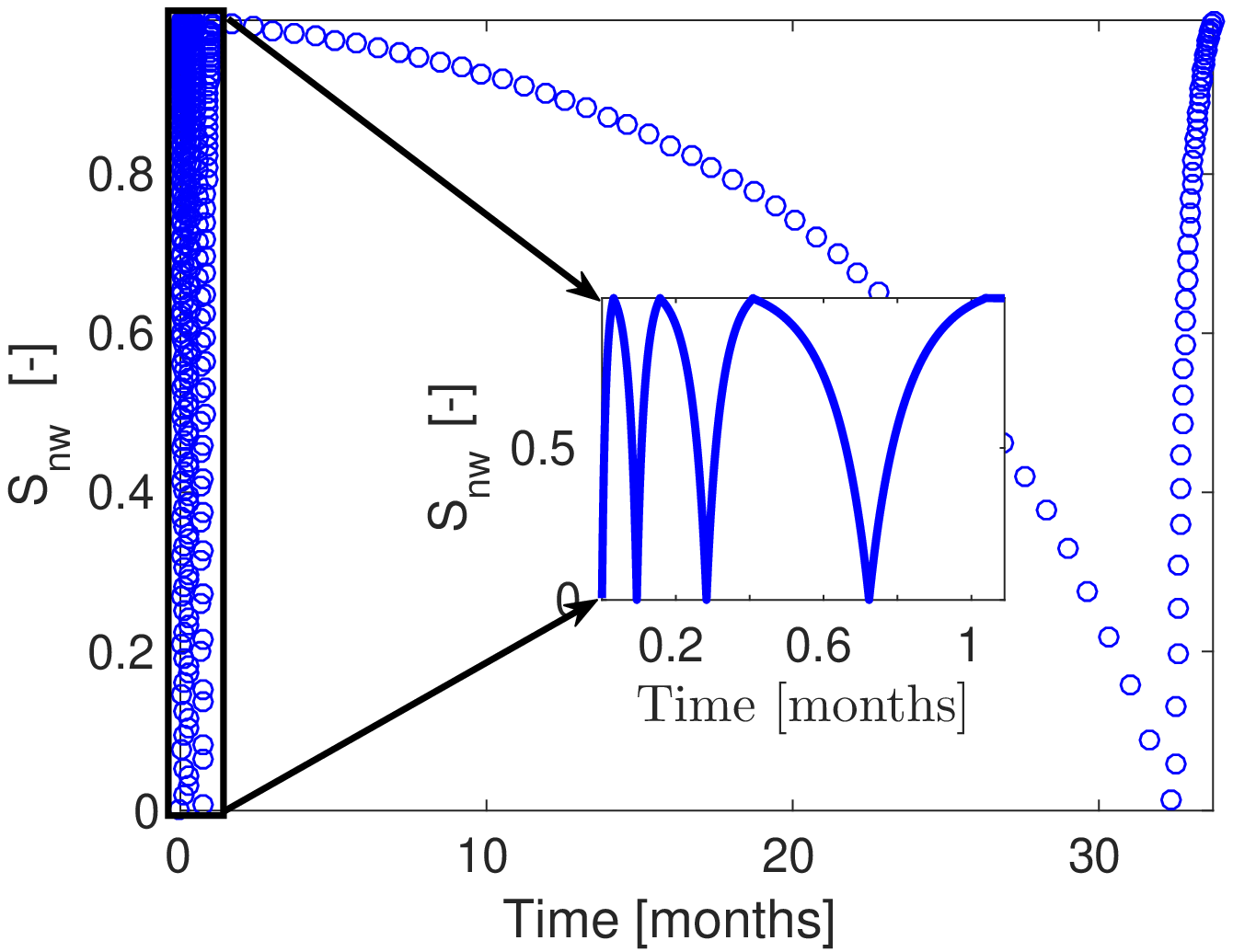}
\caption{Non-wetting fluid history paths over the four displacement cycles (a) and the averaged variable $\overline{\chi}$ which obtained by integration of the fluid history in (a) over the exposure period (b). } \label{fig:exnon}
\end{figure}

As we noticed for the uniform WA case, we also observe \pcs hysteresis induced by non-uniform WA in Figure \ref{fig4:4s}. Therefore, we can transform the \pcs curve to the temporal domain $\overline{\chi}$ in the same way as before. The temporal domain $\overline{\chi}$ is shown in Figure \ref{fig:exnon}a, and  is obtained by integrating the averaged non-wetting fluid history in Figure \ref{fig:exnon}b over exposure period of 33 months. For the sake of clarity and the same reason as the uniform WA case, we controlled the pressure drop after the third cycle so that the tubes experience prolonged exposure time in order to complete the alteration processes within few number of cycles.

\begin{figure}[t!]
\centering
\includegraphics[scale=0.65]{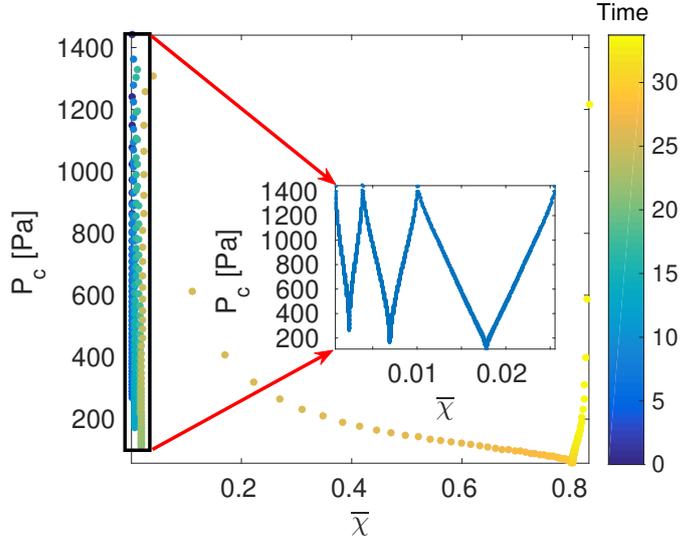}
\caption{Simulated capillary pressure curves for the case of non-uniform WA over four drainage-imibition cycles with respect to temporal domain $\overline{\chi}$. The color code shows the dynamics of capillarity during the exposure time of approximatly 33 months.}\label{fig4:4schi}
\end{figure}

The averaged variable $\overline{\chi}$ calculated from the saturation history is then used to transform the capillary pressure curve into temporal space. In Figure \ref{fig4:4schi}, we observe a unique relation between the capillary pressure and  $\overline{\chi}$ as hypothesised above. However, unlike the \pcs relation, the \pcc relation is not monotonic which rises and falls along exposure time.

\subsubsection{Dynamic capillary pressure model development}
\begin{figure}[t!]
\centering
\includegraphics[scale  =0.45]{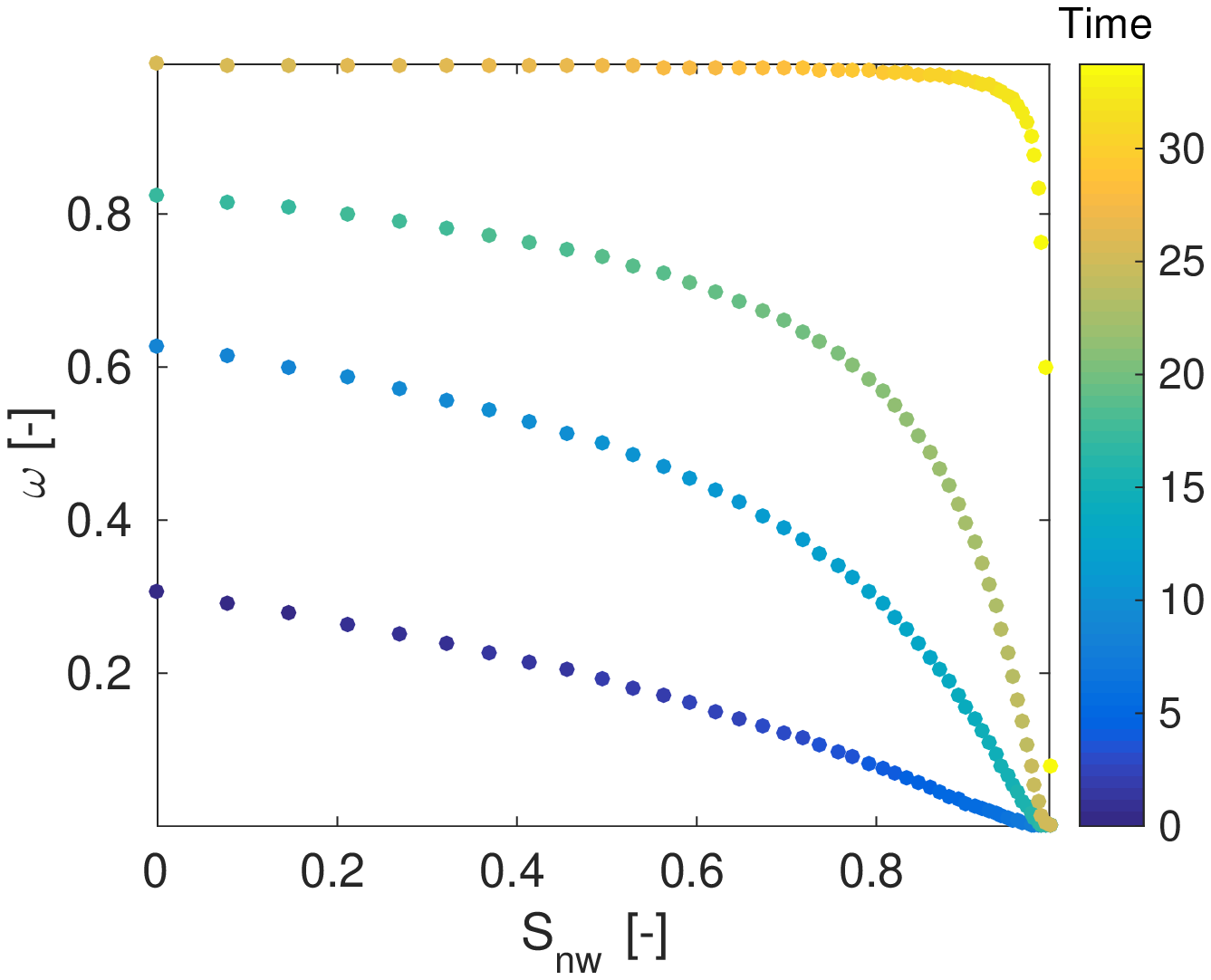}
 \includegraphics[scale =0.45]{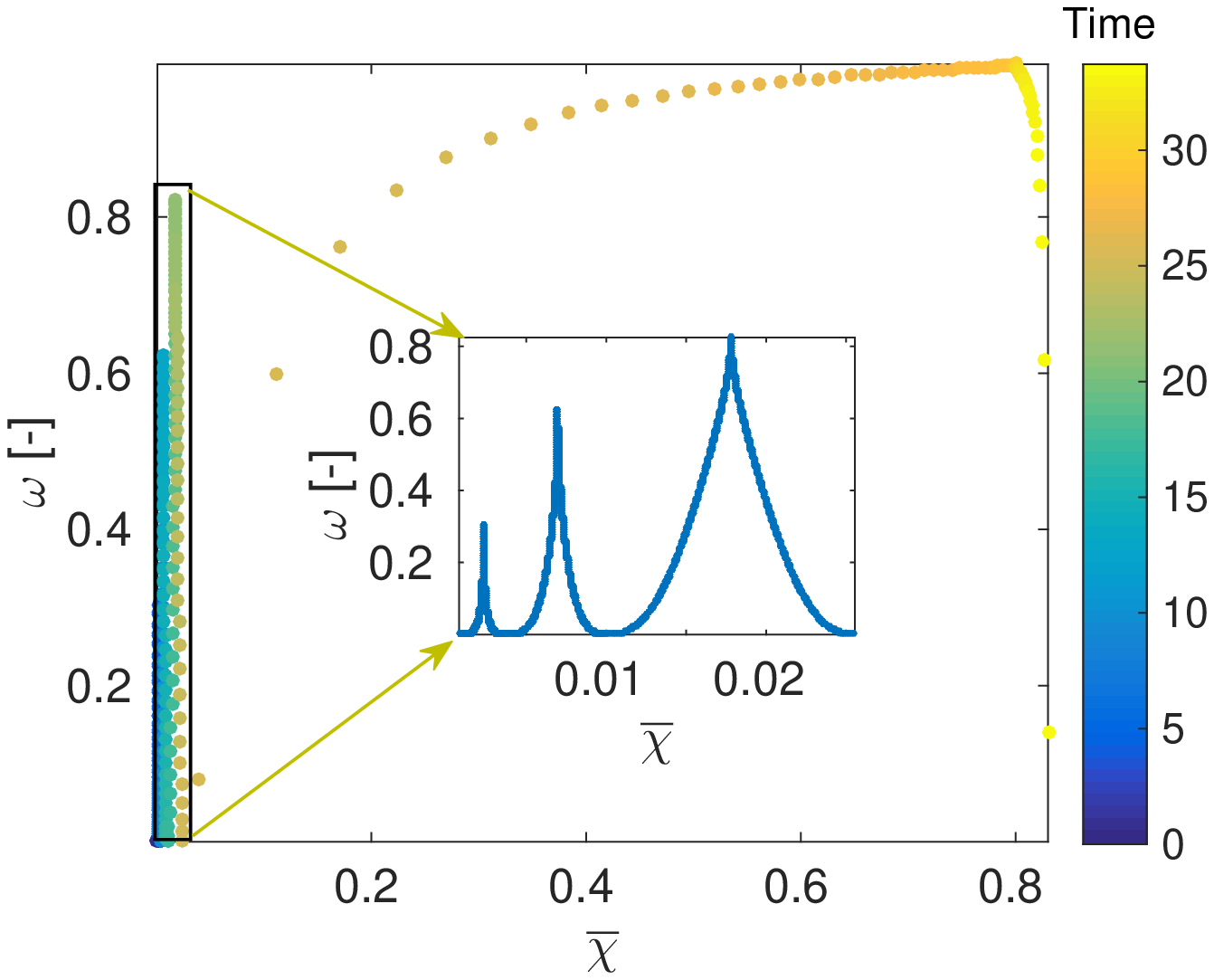}
 \caption{The dynamic coefficient obtained by applying equation (\ref{eq:dynPc2}): as a function of saturation (a) and $\overline{\chi}$ (b). Here also the data points are color-coded with exposure time in months.}\label{fig4:4dif2}
\end{figure}

Given the capillary pressure results above, the dynamic coefficient $\omega$ is calculated according to Equation (\ref{eq:dynPc2}), and plotted as a function of $S_{nw}$ and $\overline{\chi}$ in Figure \ref{fig4:4dif2}a and \ref{fig4:4dif2}b respectively. 
Unlike the uniform WA case, seen in Figure \ref{fig4:11}b, the coefficient $\omega$ is not monotonically increasing in $\overline{\chi}$, but instead it rises and falls. This makes it challenging to propose a functional relation between the dynamic coefficient $\omega$ and $\overline{\chi}$ directly as we did for the uniform WA case. 

On the other hand, $\omega$ in Figure \ref{fig4:4dif2}a has a well-behaved curvature for each drainage-imbibition cycle along the saturation history. Further, the dynamic coefficient as a function of saturation is increasing with increased exposure time. Given these insights, we proposed a model for the dynamic coefficient,
\begin{equation}\label{imbibnonuni_1}
\omega(S_{nw},\overline{\chi} )= \frac{S_{w}}{\alpha(\overline{\chi})  + S_{w}},
\end{equation}
where $S_w = 1-S_{nw}$, and $\alpha$ controls the curvature of the $\omega$-$S_{nw}$ curve for each drainage-imbibition cycle.  Since $\omega$ is increasing function of exposure time, $\alpha$ should decrease along the averaged variable $\overline{\chi}$. 

The function form of $\omega$ in (\ref{imbibnonuni_1}) is then matched with the $\omega$-$S_{nw}$ data to analyze the dynamics of $\alpha$ along $\overline{\chi}$. The obtained $\alpha$-$\overline{\chi}$ relation is decreasing   as hypothesized and in particular has a form that can be governed as,
\begin{equation}\label{alpha}
\alpha(\overline{\chi}) = \beta_2/\overline{\chi}, 
\end{equation}
where $\beta_2$ is non-dimensional fitting parameter. For this particular simulation the parameter $\beta_2$ is estimated to be 0.004 for the four of drainage-imbibition cycles.

The suggested models (\ref{imbibnonuni_1}) and (\ref{alpha}) are substituted  into the interpolation model (\ref{eq:dynPc_interp}) to give:
\begin{equation}
P_c= \frac{ \overline{\chi}S_{w}}{\beta_2  + \overline{\chi}S_{w}} \Big(P_c^{\rm st,f} - P_c^{\rm st,i} \Big)+ P_c^{\rm st,i}.
    \label{eq:dynPc_interp_re_nonuni}
\end{equation}  

\begin{figure}[t!]
\centering
\includegraphics[width=0.65\textwidth]{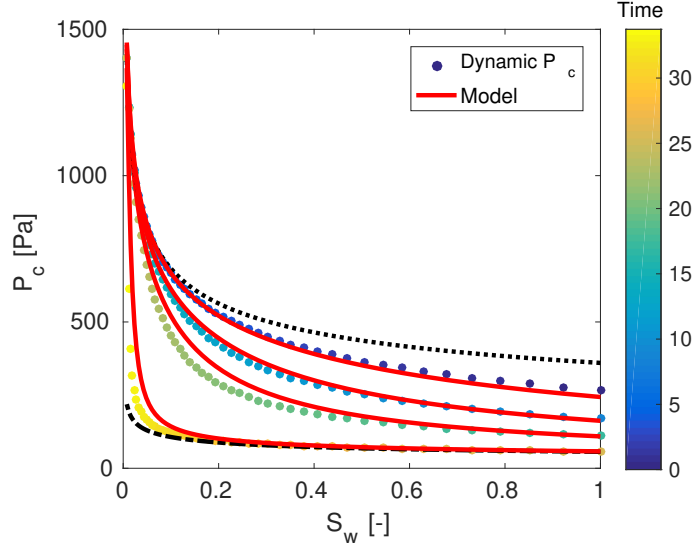}
\caption{Comparison of the dynamic model (\ref{eq:dynPc_interp_re_nonuni}) with the simulated \pcs data for the case of non-uniform wettability alteration. The data points are color-coded with exposure time in months}\label{fig:imbib}
\end{figure}

The calibrated dynamic capillary pressure model  is compared with the simulated time-dependent capillary pressure data  and depicted in Figure \ref{fig:imbib}. We observe that the proposed dynamic model agrees well with simulated dynamic capillary pressure curves. 
Thus, we have obtained a single valued single parameter  model that describes the evolution of dynamic capillarity over multiple drainage-imbibition cycles rather than using a model consisting of multiple parameters that change with each cycle (or hysteresis models). 

\subsubsection{Dynamic model applicability} 
\label{sub:dynamic_model_applicability}

\begin{figure}[t!]
\centering
\includegraphics[scale=0.45]{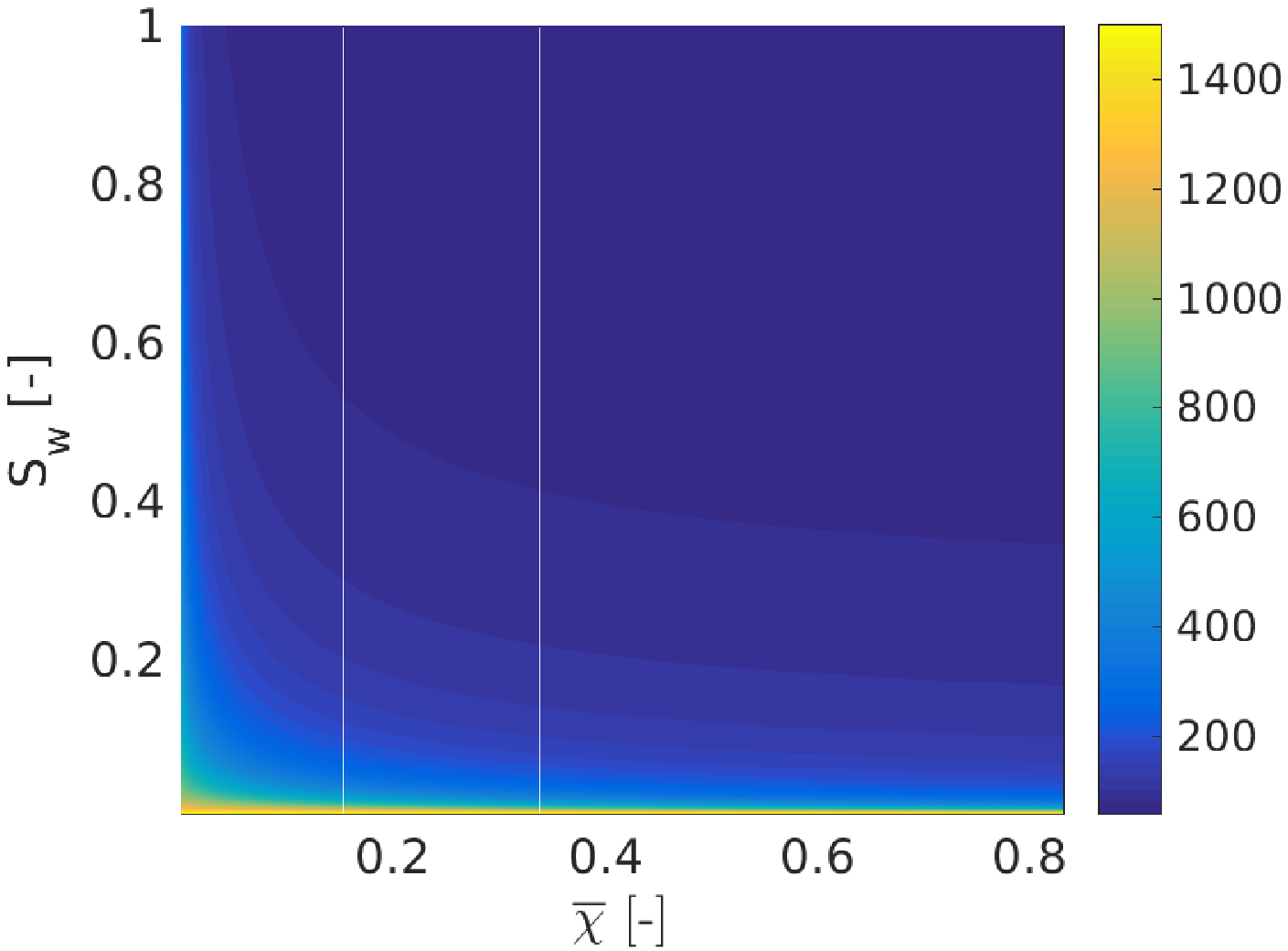}
\includegraphics[scale=0.45]{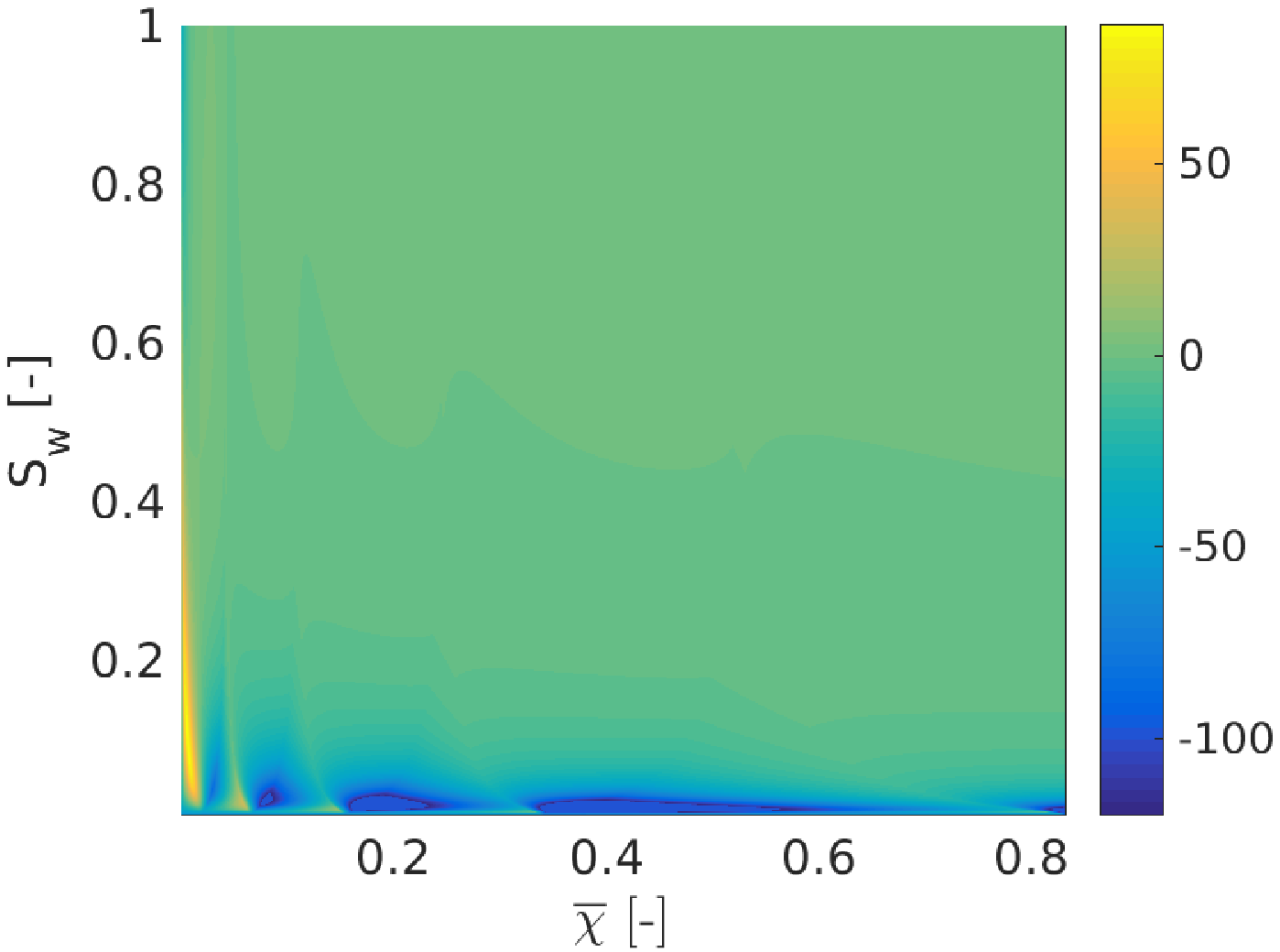}
\caption{(a) Simulated capillary pressure under non-uniform WA obtained by taking multiple paths in the $S_w\times\overline{\chi}$ domain. (b) is the difference between the dynamic capillary pressure (\ref{eq:dynPc_interp_re_nonuni}) and the \pcc data in (a).}\label{fig:dynamic_entry_non-core}
\end{figure}

As we point out in the uniform WA case, the saturation paths above are arbitrary paths out of infinitely many \pcsc paths. Keeping this in mind, we generate a data for dynamic capillary pressure  by considering possibly many paths in $S_w\times\overline{\chi}$ domain, resulting in a capillary pressure surface  shown in Figure \ref{fig:dynamic_entry_non-core}a. 

Following the paths used to generate the simulated \pcsc surface, we applied the calibrated model (\ref{eq:dynPc_interp_re_nonuni}) to predict the simulated capillary pressure data. Figure \ref{fig:dynamic_entry_non-core}b shows the actual difference between the simulated data and the model predicted data. The comparison demonstrates that a single parameter value model calibration to a single arbitrary path in $S_w\times\overline{\chi}$ domain can be applied to any possible path in the domain.  In general, this simple model is efficient to predict the the impact of  non-uniform WA at the pore scale on the capillary pressure at the macroscale for any given exposure time and saturation path.  

\begin{figure}[th!]
\centering
\includegraphics[width=0.65\textwidth]{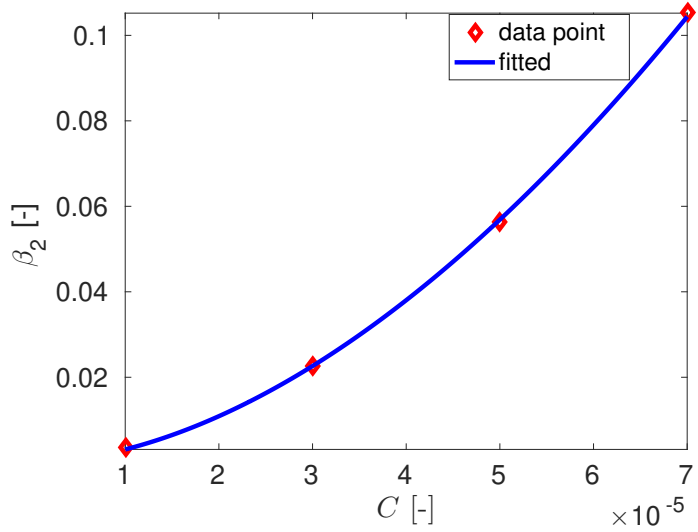}
\caption{The relation between pore-scale wettability model parameter $C$ and the dynamic coefficient parameter $\beta_2$ in Equation (\ref{eq:dynPc_interp_re_nonuni}).}\label{pararela2}
\end{figure}

Given the successful  WA induced dynamic capillary pressure model, we perform the same exercise as before where we vary the pore-scale parameter $C$, from Equation (\ref{contactGnera}), and observe its impact on the macroscale parameter $\beta_2$ in the interpolation model above. The relation between those parameters is obtained and depicted in Figure \ref{pararela2}.
A power law model, $\beta_2 = b_1C^{b_2}$ is observed between these parameters and the model parameters are estimated to be $b_1=3.3 \times 10^6$ and $b_2= 1.8$. The model for $\beta_2$ can be substituted into model (\ref{eq:dynPc_interp_re_nonuni}) to characterize the dynamics of the capillary pressure directly from the pore scale phenomenon.

To summarize, we have obtained a reliable  model to describe the impact of the pore-scale wettability alteration on capillary pressure dynamics.
\section{Discussion and Conclusions}\label{sec:5}

We have proposed a new interpolation based model for dynamic capillary pressure caused by time-dependent wettability alteration. The model has been correlated to pore-scale simulations of capillary pressure-saturation curves where a time-dependent model for contact angle change has been applied directly at the pore scale. The resulting correlation shows that  a simple interpolation-based model is highly effective at capturing WA induced dynamics in macroscale capillary pressure by a single value parameter.    In the interest of completeness, we explored other types of models to capture capillary pressure dynamics, including the mixed-wet model of Skj\ae veland et al \cite{RefSkjaeveland}. For brevity, we do not report the results of that separate study herein. We found that although other models could be calibrated with reasonable accuracy, they all involved more than one calibration parameter (up to four) that need to be adjusted in each drainage-imbibition cycle. Therefore, the single-parameter single-valued  interpolation model presented in this study is the preferred model due to its reliability. 

The proposed interpolation model is a macroscale model that allows for change in capillary pressure as a function of macroscale variables--saturation and exposure time to a WA agent. We recall that exposure time is simply the integration of saturation history over time. The model consists of three main components--two capillary pressure functions at the initial and final wetting state and a dynamic interpolation coefficient that moves from one state to the other. The initial and final capillary pressure functions can be determined \emph{a priori} from static experiments using inert fluids. In this study, the initial and final states are represented by classical Brooks-Corey functions. 
The dynamic coefficient is thus the only variable correlated to dynamic capillary pressure simulations. In this study, we have shown that the coefficient can be easily correlated to saturation and exposure time via a single parameter. 

The resulting correlation is dependent on the conditions for wettability alteration. Two conditions are explored, the first is dissolution of the agent into the wetting phase that alters the contact angle uniformly across the REV. The second is contact angle change only through direct contact with an altering agent as a separate non-wetting phase, resulting in non-uniform contact angle change throughout the REV. The differences in the two wettability alteration mechanisms changes the complexity of the resulting capillary dynamics. In the uniform case, the dynamic coefficient can be correlated to exposure time through a sorption-type model, which seems to be a natural result given the contact angle change at the pore scale is also based on a sorption model. 
 In the non-uniform case, the dynamic coefficient has a similar sorption form with increased exposure time, but now with the product of saturation and exposure time as the dynamic variable. The additional complexity is needed to draw the capillary pressure curve back to the initial wetting at low saturation (a region of the \pcs curve dominated by smaller pores where the contact angle takes longer to change).

An important result of this study is quantifying the link between the pore- and macroscale. We showed that by varying the parameter that alters the speed and extent of contact angle change in each individual pore, we could predict the resulting impact on dynamic capillary pressure. In fact, in both the uniform and non-uniform cases, there is a very simple scaling from the pore-scale and macroscale parameters. In the uniform case, the two parameters are directly proportional, while in the non-uniform case, the macroscale parameter scales with the pore-scale parameter via a power law. The implication of this result is that by knowing the mechanism that controls contact angle at the pore-scale, which can be obtained by a relatively simple batch experiment, we can quantify \emph{a priori} the macroscale dynamics without having to perform pore-scale simulations. This is an important generalization and valuable for making use of experimental data to inform macroscale constitutive functions.

The success of the interpolation model in this study is a natural development from previous studies that incorporates the interpolation model directly into reservoir simulation of wettability alteration. In those studies, the model was matched directly to macroscale data in a heuristic manner. The contribution of this study is to quantify the pore-scale underpinnings to the interpolation model through a direct and systematic manner, thus confirming the validity of this type of model for use in macroscale simulation. The insights gained in this study into capillary pressure dynamics due to wettability alteration. Further study is needed to calibrate the macroscale dynamic parameter to actual data. 

There are additional aspects that were not addressed in this study. First, is whether the effect of time-dependent wettability alteration can be quantified in a similar manner for relative permeability. For brevity, we did not explore this additional constitutive relationship in this paper. This can be explored in future work. An additional quantity of interest is residual saturation. Quantifying dynamics and stability of residual wetting and non-wetting saturation is an active area of research. The scope of this work involved a bundle-of-tubes pore-scale model, which due to its simple nature cannot model trapped phases. Therefore, a more realistic pore-network model would need to be implemented, which is the subject of ongoing work. 

The results of this study have important implications for future laboratory investigations. Previous experimental studies have shown that wettability alteration can cause a significant decrease in capillarity that occurs at the same time scale as the experiment. Our work further highlights this time-dependent effect and shows that proper quantification of the dynamics requires capillary pressure data from both inert and reactive fluid pairings. 
 In addition, there should be separate batch measurements on mineral surfaces to characterize the mechanistic model for contact angle change at the pore scale, such as the one proposed herein. Also important is whether the alteration mechanism occurs by dissolution of a WA agent or by direct contact with a non-wetting phase. Such data are important for determining the macroscale parameters needed in the interpolation-based model. 

At the reservoir scale, there is a potential for dynamics in capillary pressure to have a significant impact on fluid flow that should be accounted for during simulations. For instance, for \co storage could there be a change in wetting that could degrade the integrity of the sealing caprock and allow for seepage of \co out of the storage reservoir, and thus further investigation is warranted. This will require implementation of dynamic capillary pressure functions into reservoir simulation, with possible wettability reversal that could prove challenging for numerical consistency and robustness. Modified  simulation methods can then be used in the future to quantify the impact of capillary dynamics on reservoir processes for both hydrocarbon recovery and \co storage.

\section*{Acknowledgement}
The research is supported by CHI project funded by the CLIMIT program of the Research Council of Norway.

\bibliographystyle{plainnat}      
 

\bibliographystyle{numeric}

\end{document}